\documentclass[reprint,twocolumn,showpacs,showkeys,preprintnumbers,superscriptaddress,amsmath,amssymb,longbibliography,floatfix,aps,prb]{revtex4-1}

\usepackage{graphicx}% Include figure files
\usepackage{bm}% bold math

\usepackage{placeins}
\usepackage{graphics}
\usepackage{color}
\usepackage{hyperref}
\usepackage{multirow}
\usepackage{blindtext}
\usepackage{upgreek}
\usepackage{braket}

\begin{document}

%\preprint{APS/123-QED}

\title{Graphene on two-dimensional hexagonal BN, AlN, and GaN:\\ Electronic, spin-orbit, and spin relaxation properties}

\author{Klaus Zollner}
\email{klaus.zollner@physik.uni-regensburg.de}
\affiliation{Institute for Theoretical Physics, University of Regensburg, 93040 Regensburg, Germany}
\author{Aron W. Cummings}
\affiliation{Catalan Institute of Nanoscience and Nanotechnology (ICN2), CSIC and The Barcelona Institute of Science and Technology, Campus UAB, Bellaterra, 08193 Barcelona, Spain}
\author{Stephan Roche}
\affiliation{Catalan Institute of Nanoscience and Nanotechnology (ICN2), CSIC and The Barcelona Institute of Science and Technology, Campus UAB, Bellaterra, 08193 Barcelona, Spain}
\affiliation{ICREA - Instituci\'{o} Catalana de Recerca i Estudis Avan\c{c}ats, 08010 Barcelona, Spain}
\author{Jaroslav Fabian}
\affiliation{Institute for Theoretical Physics, University of Regensburg, 93040 Regensburg, Germany}

\begin{abstract}
We investigate the electronic band structure of graphene on a series of two-dimensional hexagonal nitride insulators hXN, X = B, Al, and Ga, with first principles calculations. A symmetry-based model Hamiltonian is employed to extract orbital parameters and spin-orbit coupling (SOC) from the low-energy Dirac bands of the proximitized graphene. While commensurate
hBN induces a staggered potential of about 10~meV into the 
Dirac band structure, less lattice-matched hAlN and hGaN disrupt the Dirac point much less, giving a staggered gap below 100 $\upmu$eV. Proximitized intrinsic SOC surprisingly does not increase much above the pristine graphene value of 12 $\upmu$eV; it stays in the window of (1-16) $\upmu$eV, depending strongly on stacking. However, Rashba SOC increases sharply when increasing the atomic number of the boron group, with calculated maximal values of 8, 15, and 65 $\upmu$eV for B, Al, and Ga-based nitrides, respectively. The individual Rashba couplings also depend strongly on stacking, vanishing in symmetrically-sandwiched structures, and can be tuned by a transverse electric field. The extracted spin-orbit parameters were used as input for spin transport simulations based on Chebyshev expansion of the time-evolution of the spin expectation values,
yielding interesting predictions for the electron spin relaxation.
Spin lifetime magnitudes and anisotropies depend strongly on the specific (hXN)/graphene/hXN system, 
and they can be efficiently tuned by an applied external electric field as well as the carrier density in the graphene layer. A particularly interesting case for experiments is graphene/hGaN, in which the giant Rashba coupling is predicted to induce spin lifetimes of 1-10 ns, short enough to dominate over other mechanisms, and lead to the same spin relaxation anisotropy as that observed in conventional semiconductor heterostructures: 50\%, meaning that out-of-plane spins relax twice as fast as in-plane spins.  
\end{abstract}

\maketitle

%------------------------------------------------------------
\section{Introduction}
%------------------------------------------------------------

Hexagonal boron nitride (hBN) has become one of the most important insulator materials for electronics and spintronics.
In the two-dimensional (2D) few-layer limit it is often used as a tunnel barrier to overcome the conductivity mismatch in spin injection devices, or as an encapsulation material to protect other 2D materials from degradation.
The current generation of graphene-based spintronic devices also relies on the excellent properties of hBN as a substrate material, leading to outstanding spin and charge transport properties \cite{Banszerus2015:SA, Petrone2012:NL, Calado2014:APL, Drogeler2016:NL, Zollner2019:PRB, Roche2015:2DM,Kamalakar2014:APL,Gurram2017:2DM,Guimaraes2014:PRL,Singh2016:APL,Drogeler2016:NL, Drogeler2017:PSS, Zomer2012:PRB, Ingla-Aynes2015:PRB, Drogeler2014:NL, Dean2012:SSC, Dean2010:NN, Wang2017:RSC}, 
which are highly important for the realization of spin logic devices \cite{Gurram2016:PRB, Mayorov2011:NL,Gurram2017:NC, Gurram2018:PRB, Gurram2017:2DM,Han2014:NN, Kamalakar2014:APL, Britnell2012:Sc, Wang2017:MTP, Fabian2007:APS, Zutic2004:RMP, Lin2013:NL,Lin2014:ACS,Luo2017:NL,Wen2016:PRA,Zutic2006:IBM,Behin2010:NN}.
However, there are other layered hexagonal nitride semiconductors/insulators that may prove useful for electronic and spintronic applications. In particular, first-principles calculations have demonstrated the chemical stability of hexagonal aluminum nitride (hAlN) and hexagonal gallium nitride (hGaN) \cite{Kecik2015:PRB, Bacaksiz2015:PRB, Zhuang2013:PRB, Sahin2009:PRB, Vahedi2015:SM, Peng2013:RCS, Freeman2006:PRL}, as well as revealing their mechanical and optical properties.

Bulk AlN is an important material\cite{Beshkova2020:V}. It is used as a high-quality substrate for the growth of topological insulators \cite{Tsipas2014:ACS, Freitas2016:APL} and transition-metal dichalcogenides (TMDCs) \cite{Xenogiannopoulou2015:Nano, Ouyang2016:PCCP}.
To grow strain-free AlN on a sapphire substrate, graphene can 
be used as a buffer layer\cite{Qi2018:JACS}.
The two-dimensional counterpart to bulk AlN is also becoming more important. There is strong experimental \cite{Tsipas2013:APL, Mansurov2015:JCG} and theoretical \cite{Bacaksiz2015:PRB, Zhuang2013:PRB, Sahin2009:PRB} evidence of graphite-like hAlN, which can be thinned down to the monolayer limit. Graphene/hAlN heterostructures have already been considered theoretically \cite{DosSantos2016:Nano} and
just recently have been grown \cite{Wang2019:AM}.

Similarly, GaN is an important semiconductor material for technological applications, especially in the wurtzite form with a direct band gap in the ultraviolet range \cite{Nakamura2015:RMP, Onen2016:PRB}. Monolayer hGaN is therefore a potentially important material for ultra-compacted electronics and optics \cite{Kecik2018:APL}.
Interestingly, graphene also seems to play a major role in the growth of hGaN \cite{AlBalushi2016:NM}.
Graphene on hGaN was considered from first-principles \cite{Deng2019:RSC} as a tool to control the Schottky barrier and contact type by strain engineering. 

There are already several fields now in which hAlN and hGaN show promising results.
For example, TMDC/hAlN or blue phosphorene/hGaN heterostructures are predicted to be important visible-light-driven photocatalysts for water splitting \cite{Liao2014:JPCC, Ren2019:RSC}.
Excitonic effects have also been studied in hAlN and hGaN \cite{Vahedi2015:SM, Prete2019:arxiv}, with 
absorption spectra dominated by strongly bound excitons. 
Both hAlN and hGaN are gaining interest for opto-electronic applications \cite{Kecik2018:APL} due to their two-dimensional nature. Stacks of hAlN and hGaN \cite{Onen2017:PRB} also show novel electronic and optical properties.

From the theoretical perspective, the properties of hAlN and hGaN are not yet well studied. 
Monolayer hAlN has an indirect band gap ranging between \mbox{$2.9$--$5.8$~eV} \cite{DosSantos2016:Nano, Prete2017:APL}, which is strain dependent \cite{Liu2014:CMS, Kecik2015:PRB}. 
The band gap also depends on the number of hAlN layers \cite{Kecik2015:PRB}. 
Monolayer hGaN has an indirect band gap ranging between $2.1$--$5.5$~eV, which is also tunable by strain \cite{Onen2016:PRB, Sun2017:APL, Kecik2018:APL, Prete2017:APL}. 
$G_0W_0$-calculations provide band gaps of hAlN (5.8~eV \cite{Prete2017:APL}) and hGaN (4.55~eV \cite{Onen2016:PRB}) comparable to the theoretical and experimental values for hBN (6.0~eV \cite{Ferreira2019:OSA}). It then seems natural to consider  hGaN and hAlN as alternatives to hBN in transport experiments, even though few-layer and monolayer materials are difficult to synthesize at the moment. 

Layered hBN is a perfect substrate for graphene spintronics, allowing for giant mobilities in graphene/hBN structures \cite{Roche2015:2DM}, ultralong spin lifetimes \cite{Banszerus2015:SA,Drogeler2014:NL,Drogeler2016:NL,Yang2011:PRL,Drogeler2017:PSS}, for extracting spin-orbit gaps in bilayer graphene \cite{Banszerus2020:PRL}, or for revealing surprisingly strong spin-orbit anisotropy in bilayer graphene \cite{Xu2018:PRL,Leutenantsmeyer2018:PRL}.  On the other hand, hBN induces only weak SOC in graphene \cite{Zollner2019:PRB}, limiting its potential for investigating proximity effects. Would hAlN and hGAN provide the same protection for the electron spins in graphene as hBN? What is the expected spin relaxation if graphene is stacked with the two alternative insulators? In the 
absence of experimental studies, theoretical answers to these questions are particularly important. 

Here, we use first principles calculations to study graphene on monolayers of hBN, hAlN, and hGaN.
We investigate the proximity-induced SOC in graphene by fitting a low energy model Hamiltonian to the Dirac bands. In particular, we find that the lattice-matched hBN induces a strong sublattice asymmetry, resulting in an orbital gap on the order of 20~meV, while there is an averaging effect for the sublattice asymmetry in the less lattice-matched hAlN and hGaN substrates, resulting in orbital gaps below 200~$\upmu$eV. The intrinsic SOC parameters stay below 15~$\upmu$eV for all hXN substrates, where X = B, Al, and Ga, while the Rashba 
SOC increases from a few to tens of $\upmu$eV with ascending atomic number of the boron group. 
By varying the interlayer distance between graphene and a given hXN substrate, we not only find the energetically most favorable geometry, but also show that the orbital gap as well as proximity-induced SOC  parameters [especially Rashba and pseudospin-inversion asymmetry (PIA) ones] are highly tunable by the van der Waals gap between the layers (more than 100\% by tuning the interlayer distance by only 10\%). 
In addition, we show that an external transverse electric field can be used to tune the SOC of the proximitized graphene. While the intrinsic couplings can be tuned between 5 and 20~$\upmu$eV, the Rashba and PIA couplings can reach up to a few hundred $\upmu$eV, within our considered field limits of $\pm 3$~V/nm. 
By encapsulating graphene between two hXN layers, the induced SOC results from an interplay of both layers (hBN induces a sizeable orbital gap, 
while additionally, e. g., hGaN provides a large Rashba and PIA SOC) allowing for the customization of spin transport in graphene via layer engineering.  

Combining the model Hamiltonian and fitted parameters, we then perform spin transport simulations to gain insight into spin relaxation in (hXN)/graphene/hXN heterostructures. 
In all studied systems, we find large spin lifetimes in the nanosecond range and above, reaching up to a few seconds in encapsulated structures. Depending on the heterostructure, strong electron/hole asymmetry and giant anisotropy in the spin lifetimes can be observed, which are experimentally testable fingerprints of our results.
For example, the weak intrinsic combined with the strong Rashba SOC in graphene/hGaN results in spin lifetimes between 1 and 10~ns and a nearly constant spin relaxation anisotropy of 50\% for all carrier densities. Adding a hBN capping layer, meanwhile, introduces a sizeable orbital band gap in the Dirac spectrum, resulting in giant out-of-plane spin lifetimes and anisotropies near the charge neutrality point.

%-------------------------------------------------------------
\section{Computational Details and Geometry}
%-------------------------------------------------------------
%------------------------------------------------------------------------
\begin{figure*}[htb]
 \includegraphics[width=.99\textwidth]{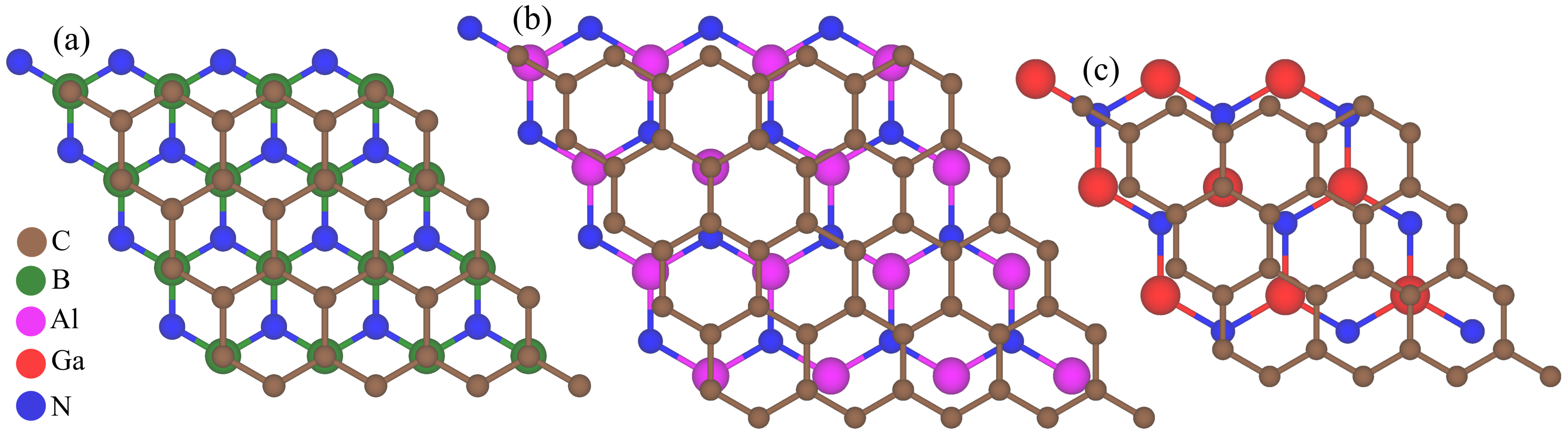}
 \caption{(Color online) Top view of the supercell geometry of 
 (a) graphene on hBN ($4 \times 4$), (b) graphene on hAlN, and (c) graphene on hGaN.
 Different colors correspond to different atom types.
 }\label{Fig:AlN_GaN_Structures}
\end{figure*}
%------------------------------------------------------------------------

%------------------------------------------------------------------------
\begin{figure*}[htb]
 \includegraphics[width=.99\textwidth]{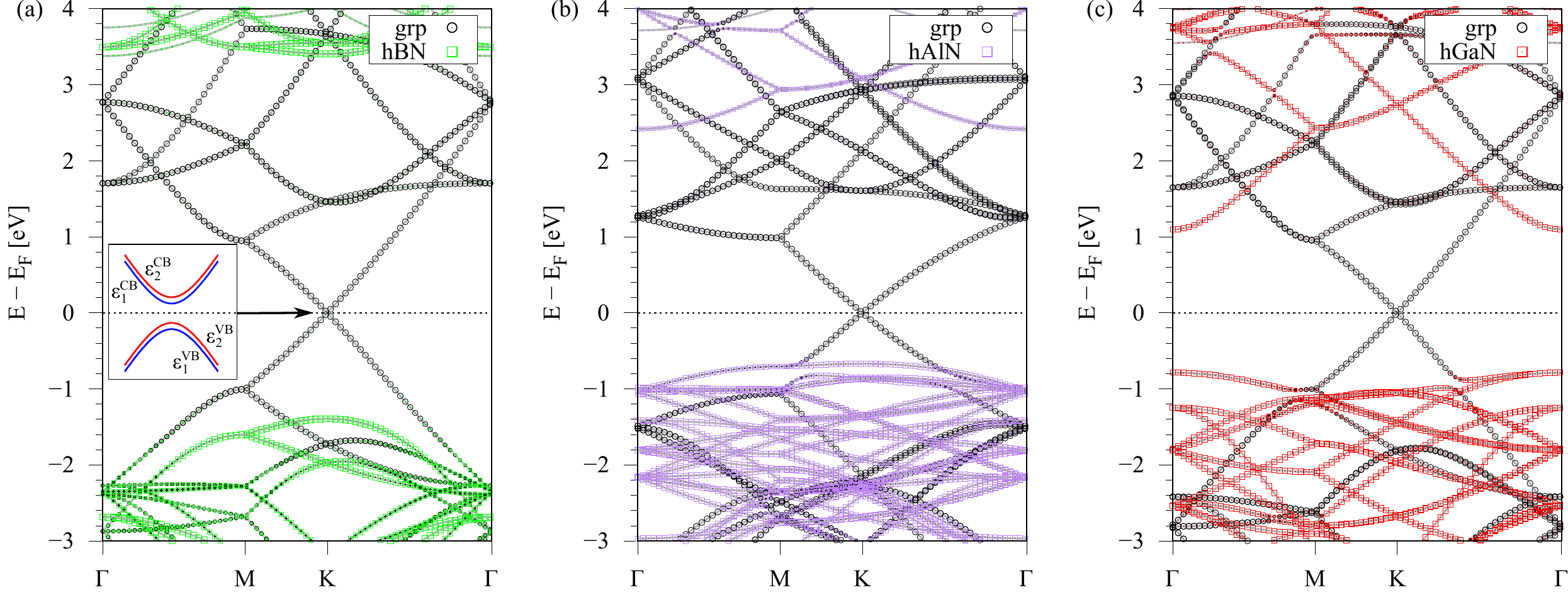}
 \caption{(Color online) Calculated band structures of graphene on (a) hBN, (b) hAlN, and (c) hGaN. Bands corresponding to different layers (graphene, hBN, hAlN, hGaN) are plotted in different colors (black, green, purple, red). 
 The inset in (a) shows a sketch of the low energy dispersion close to the K point.
  Due to the presence of the substrate, graphene's low energy bands are split into four states $\varepsilon_{1/2}^{\textrm{CB/VB}}$, with a band gap.}
  \label{Fig:bands}
\end{figure*}
%------------------------------------------------------------------------

In Fig.~\ref{Fig:AlN_GaN_Structures} we show the supercell geometries
of the graphene/hBN, graphene/hAlN, and graphene/hGaN heterostructures.
The lattice constant of graphene \cite{Neto2009:RMP} is $a = 2.46$~\AA, the one of hBN \cite{Catellani1987:PRB} is $a = 2.504$~\AA, the one of hAlN \cite{Zhuang2013:PRB} is $a = 3.12$~\AA, 
and the one of hGaN \cite{Zhuang2013:PRB} is $a = 3.25$~\AA,  which we adapt slightly (straining the lattices by less than 2\%) for the periodic first-principles calculations. 
For the graphene/hBN structure, we consider $4\times 4$ ($5\times 5$) supercells of graphene and hBN, both with a lattice constant of $a = 2.4820$~\AA, and we use the most energetically favorable stacking for these commensurate lattices \cite{Zollner2019:PRB}.
In the case of graphene on hGaN, we place a $4\times 4$ graphene unit cell, with a compressed lattice constant of $a = 2.4486$~\AA, on a $3\times 3$ hexagonal hGaN unit cell, with a stretched lattice constant of $a = 3.2649$~\AA~ \cite{Zhuang2013:PRB}. 
For graphene on hAlN, we place a $5\times 5$ graphene unit cell, with a stretched lattice constant of $a = 2.4816$~\AA, on a $4\times 4$ hexagonal hAlN unit cell, with a compressed lattice constant of $a = 3.102$~\AA~ \cite{Zhuang2013:PRB, Bacaksiz2015:PRB,Tsipas2013:APL}.

First-principles calculations are performed with a full potential linearized augmented plane wave (FLAPW) code based on density functional theory (DFT), as implemented in WIEN2k \cite{Wien2k}. Exchange-correlation effects are treated with the generalized-gradient approximation (GGA) \cite{Perdew1996:PRL}, including a dispersion correction \cite{Grimme2010:JCP}. 

For graphene/hGaN, we use a $k$-point grid of $18\times 18\times 1$, and the values of the Muffin-tin radii are $R_{\textrm{C}}=1.33$ for C atoms, $R_{\textrm{Ga}}=1.90$ for Ga atoms, and $R_{\textrm{N}}=1.64$ for N atoms, with the plane wave cutoff parameter $RK_{\textrm{MAX}}=4.7$.
For graphene/hAlN, we use a $k$-point grid of $12\times 12\times 1$, and the Muffin-tin radii are $R_{\textrm{C}}=1.35$ for C atoms, $R_{\textrm{Al}}=1.72$ for Al atoms, and $R_{\textrm{N}}=1.64$ for N atoms, with the plane wave cutoff parameter $RK_{\textrm{MAX}}=4.2$. 
We consider only the stacking configurations shown in Fig.~\ref{Fig:AlN_GaN_Structures}.
For these two heterostructures we vary the distance between the graphene and hXN layers to find the lowest energy separation. When we then consider graphene encapsulated in hAlN or hGaN, we use these interlayer distances. 
For the encapsulated structures, a second hAlN (hGaN) layer is added on top of the graphene/hAlN (graphene/hGaN) stack, and we use an AA$^{\prime}$ stacking of the two hAlN (hGaN) layers with graphene sandwiched between.
To avoid interactions between periodic images of our slab geometries, we add a vacuum of at least $24$~\AA~in the $z$ direction for all structures we consider.

Graphene/hBN heterostructures were already extensively discussed in Ref.\ \citenum{Zollner2019:PRB}.
For purposes of comparison, we consider here $4\times 4$ ($5\times 5$) graphene/hBN supercells, using a $k$-point grid of $18\times 18\times 1$ ($12\times 12\times 1$) and an interlayer distance of $d=3.48$~\AA. 
The Muffin-tin radii are $R_{\textrm{C}}=1.35$ for C atoms,
$R_{\textrm{B}}=1.28$ for B atoms, and $R_{\textrm{N}}=1.41$ for N atoms, 
with the plane wave cutoff parameter $RK_{\textrm{MAX}}=4.7$ ($4.2$).

We also consider asymmetric hBN/graphene/hGaN and hBN/graphene/hAlN sandwich structures.
In these cases, we take the geometries for graphene/hAlN and graphene/hGaN with their energetically most favorable interlayer distances and place another hBN layer above graphene at a distance of $d=3.48$~\AA. The in-plane lattice constant of hBN is adapted to that of graphene.
For the sandwich structure including hGaN (hAlN), we use a $k$-point grid of $18\times 18\times 1$ ($12\times 12\times 1$). The Muffin-tin radii are $R_{\textrm{C}}=1.33$ ($1.35$) for C atoms, $R_{\textrm{B}}=1.26$ ($1.28$) for B atoms, $R_{\textrm{N}}=1.39$ ($1.41$) for N atoms, and $R_{\textrm{Ga}}=1.90$ ($R_{\textrm{Al}}=1.72$) for Ga (Al) atoms, with the plane wave cutoff parameter $RK_{\textrm{MAX}}=4.7$ ($4.2$).

%------------------------------------------------------------
\section{Model Hamiltonian}
%------------------------------------------------------------

The band structure of proximitized graphene can be modeled by symmetry-derived Hamiltonians \cite{Kochan2017:PRB}. For graphene in
heterostructues with $C_{3v}$ symmetry, the effective low energy  
Hamiltonian is
\begin{flalign}
\label{Eq:Hamiltonian}
&\mathcal{H} = \mathcal{H}_{0}+\mathcal{H}_{\Delta}+\mathcal{H}_{\textrm{I}}+\mathcal{H}_{\textrm{R}}+\mathcal{H}_{\textrm{PIA}},\\
&\mathcal{H}_{0} = \hbar v_{\textrm{F}}(\tau k_x \sigma_x - k_y \sigma_y)\otimes s_0, \\
&\mathcal{H}_{\Delta} =\Delta \sigma_z \otimes s_0,\\
&\mathcal{H}_{\textrm{I}} = \tau (\lambda_{\textrm{I}}^\textrm{A} \sigma_{+}+\lambda_{\textrm{I}}^\textrm{B} \sigma_{-})\otimes s_z,\\
&\mathcal{H}_{\textrm{R}} = -\lambda_{\textrm{R}}(\tau \sigma_x \otimes s_y + \sigma_y \otimes s_x),\\
&\mathcal{H}_{\textrm{PIA}} = a(\lambda_{\textrm{PIA}}^\textrm{A} \sigma_{+}-\lambda_{\textrm{PIA}}^\textrm{B} 
\sigma_{-})\otimes (k_x s_y - k_y s_x). 
\end{flalign}
Here $v_{\textrm{F}}$ is the Fermi velocity and the in-plane wave vector 
components $k_x$ and $k_y$ are measured from $\pm$K, corresponding to the valley index $\tau = \pm 1$.
The Pauli spin matrices are $s_i$, acting on spin space ($\uparrow, \downarrow$), and $\sigma_i$ are pseudospin matrices, acting on sublattice space (C$_\textrm{A}$, C$_\textrm{B}$), with $i = \{ 0,x,y,z \}$ and $\sigma_{\pm} = \frac{1}{2}(\sigma_z \pm \sigma_0)$.
The lattice constant of graphene is $a$ and the staggered sublattice potential gap is $\Delta$.
The parameters $\lambda_{\textrm{I}}^\textrm{A}$ and $\lambda_{\textrm{I}}^\textrm{B}$ describe the sublattice-resolved intrinsic SOC, $\lambda_{\textrm{R}}$ stands for the Rashba SOC, and $\lambda_{\textrm{PIA}}^\textrm{A}$ and $\lambda_{\textrm{PIA}}^\textrm{B}$ are for the sublattice-resolved pseudospin-inversion asymmetry (PIA) SOC. 
The basis states are $\ket{\Psi_{\textrm{A}}, \uparrow}$, $\ket{\Psi_{\textrm{A}}, \downarrow}$, $\ket{\Psi_{\textrm{B}}, \uparrow}$, and $\ket{\Psi_{\textrm{B}}, \downarrow}$, resulting in four eigenvalues $\varepsilon_{1/2}^{\textrm{CB/VB}}$.

%------------------------------------------------------------
\section{First-principles and Fit Results}
%------------------------------------------------------------

In Fig.~\ref{Fig:bands}, we show the calculated band structures of 
graphene on hBN, hAlN, and hGaN. 
We find that the Dirac bands are always preserved and are located inside 
the band gap of the substrate. The low energy bands of graphene
are split into four states $\varepsilon_{1/2}^{\textrm{CB/VB}}$ with a band gap, as shown in the inset in Fig.~\ref{Fig:bands}(a), which is caused by the pseudospin symmetry breaking and the proximity-induced SOC.

%------------------------------------------------------------
\subsection{Low energy bands}
%------------------------------------------------------------

%------------------------------------------------------------------------
\begin{figure}[htb]
 \includegraphics[width=.99\columnwidth]{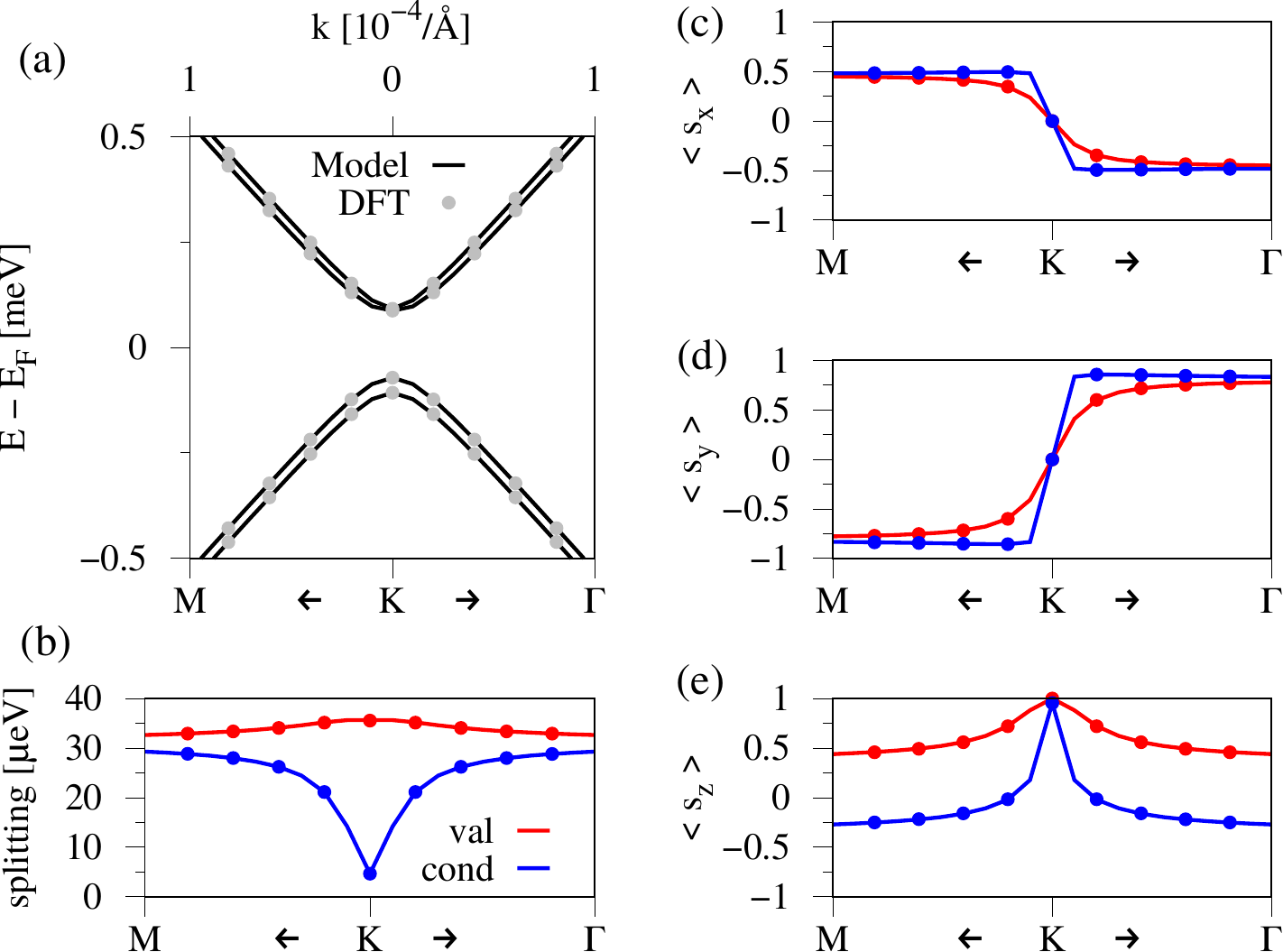}
 \caption{(Color online) Calculated low energy band properties of graphene on hAlN in the vicinity of the K point with an interlayer distance of $3.48$~\AA. 
 (a) First-principles band structure (symbols) with a fit to the model Hamiltonian (solid line). 
 (b) The splitting of conduction band $\Delta\textrm{E}_{\textrm{CB}}$ (blue) and valence band $\Delta\textrm{E}_{\textrm{VB}}$ (red) close to the K point and calculated model results. 
 (c)-(e) The spin expectation values of the bands $\varepsilon_{2}^{\textrm{VB}}$ and $\varepsilon_{1}^{\textrm{CB}}$ and comparison to the model results. 
 }\label{Fig:Fit_grp_AlN}
\end{figure}
%------------------------------------------------------------------------

%------------------------------------------------------------------------
\begin{figure}[htb]
 \includegraphics[width=.99\columnwidth]{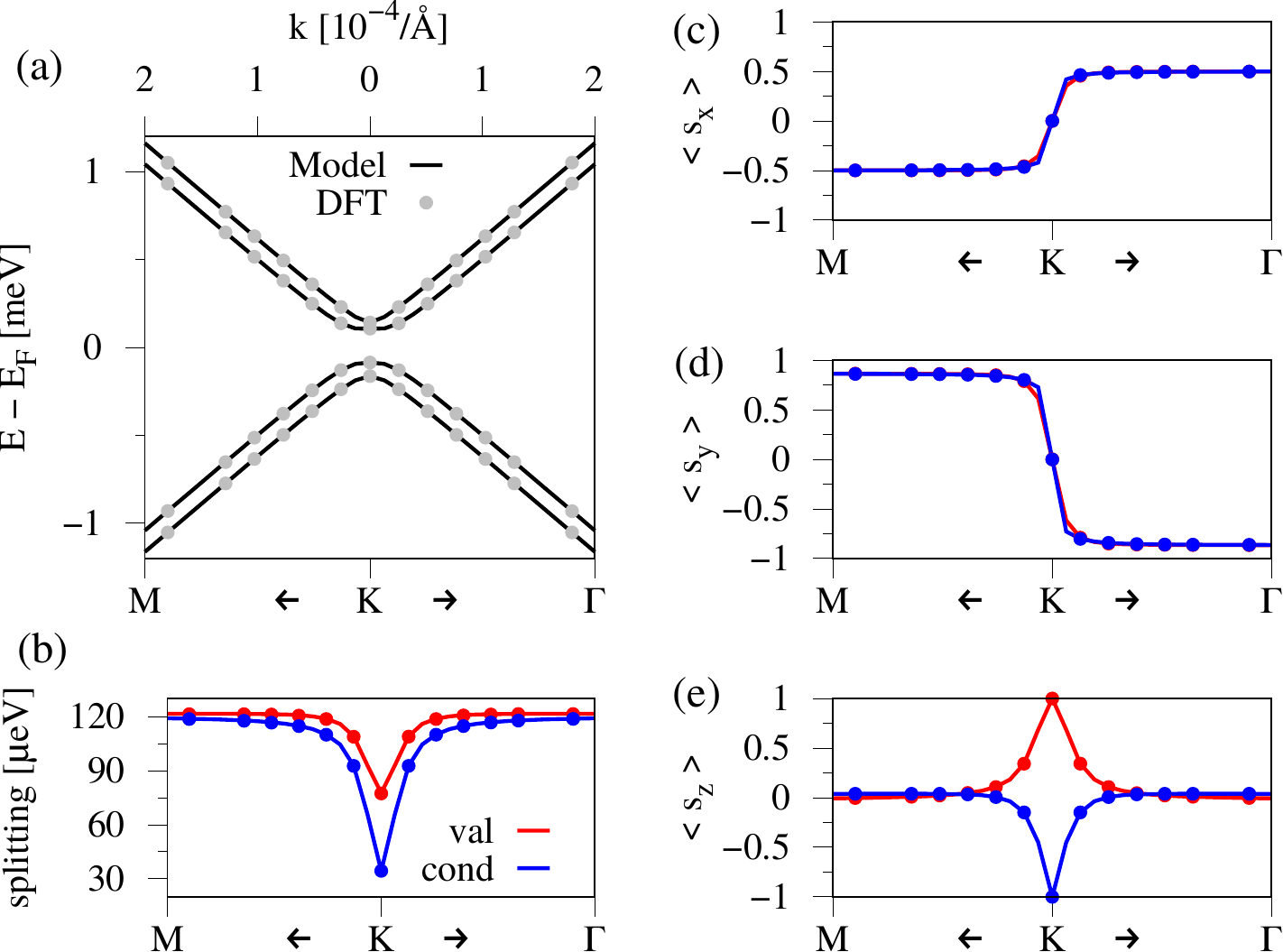}
 \caption{(Color online) Calculated low energy band properties of graphene on hGaN in the vicinity of the K point with an interlayer distance of $3.48$~\AA.
 (a) First-principles band structure (symbols) with a fit to the model Hamiltonian (solid line). 
 (b) The splitting of conduction band $\Delta\textrm{E}_{\textrm{CB}}$ (blue) and valence band $\Delta\textrm{E}_{\textrm{VB}}$ (red) close to the K point and calculated model results. 
 (c)-(e) The spin expectation values of the bands $\varepsilon_{2}^{\textrm{VB}}$ and $\varepsilon_{1}^{\textrm{CB}}$ and comparison to the model results.  
 }\label{Fig:Fit_grp_GaN}
\end{figure}
%------------------------------------------------------------------------

In Fig.~\ref{Fig:Fit_grp_AlN} we show the low energy band properties
of the graphene/hAlN stack. We find perfect agreement of the band structure, band splittings and spin expectation values with the first-principles data, in the vicinity of the K point.
In Fig.~\ref{Fig:Fit_grp_AlN}(a) we show a zoom to the low energy bands with a fit to our low energy model Hamiltonian.
The splittings, shown in Fig.~\ref{Fig:Fit_grp_AlN}(b), are around $30~\upmu$eV near the K point, and they are of the same order as for graphene on hBN \cite{Zollner2019:PRB}. 
The $s_x$ and $s_y$ spin expectation values show a clear signature of Rashba SOC, while the $s_z$ expectation value sharply decays away from the K point, see Fig.~\ref{Fig:Fit_grp_AlN}(c-e). 
In Fig.~\ref{Fig:Fit_grp_GaN} we show the low energy band properties
of the graphene/hGaN stack. The overall results are very similar, except that the band splittings are much larger, around $120~\upmu$eV near the K point.
The model is also in perfect agreement with the calculated bands.

%-----------------------------------------------------------------
\begin{table*}[htb]
\caption{\label{tab:fit} Summary of the fit parameters of Hamiltonian $\mathcal{H}$, for graphene/hBN, graphene/hGaN, and graphene/hAlN systems with interlayer distances of $3.48$~\AA. 
We have artificially turned off SOC on N, Al, Ga, or C atoms, to resolve the contributions of individual atoms to the band splittings.
The fit parameters are the Fermi velocity $v_{\textrm{F}}$, gap parameter $\Delta$, Rashba SOC parameter $\lambda_{\textrm{R}}$, intrinsic SOC parameters $\lambda_{\textrm{I}}^\textrm{A}$ and $\lambda_{\textrm{I}}^\textrm{B}$, and PIA SOC parameters  $\lambda_{\textrm{PIA}}^\textrm{A}$ and $\lambda_{\textrm{PIA}}^\textrm{B}$ for sublattices A and B. }
\begin{ruledtabular}
\begin{tabular}{l  l  c c  c  c   c  c  c c }
System & SOC on & $v_{\textrm{F}} [10^5 \textrm{m}/\textrm{s}$] & $\Delta$~[$\upmu$eV]& $\lambda_{\textrm{R}}~[\upmu$eV] & $\lambda_{\textrm{I}}^\textrm{A}~[\upmu$eV] &$\lambda_{\textrm{I}}^\textrm{B}~[\upmu$eV] & $\lambda_{\textrm{PIA}}^\textrm{A}~[\upmu$eV] & $\lambda_{\textrm{PIA}}^\textrm{B}~[\upmu$eV] \\
\hline
grp/hGaN & C, Ga, N & 8.328 & 94.9 & -60.3 & -12.5 & -9.2 & -117.5 & -117.5 \\
grp/hGaN\footnotemark[1] & C, Ga, N & 8.330 & 48.3 & -60.4 & -12.5 & -9.2 & -117.1 & -117.1 \\
grp/hGaN & Ga, N & 8.329 & 96.3 & -64.9 & -1.7 & 1.6 & -111.1 & -111.1 \\
grp/hGaN & C & 8.329 & 94.9 & 4.6 & -10.8 & -10.8 & 0 & 0 \\
hGaN/grp/hGaN & C, Ga, N & 8.272 & 27.6 & 0.1 & -10.6 & -10.6 & 0 & 0\\
\hline
grp/hAlN & C, Al, N & 8.172 & 87.1 & 14.7 & -4.6 & -15.6 & 70.3 & 70.3 \\
grp/hAlN\footnotemark[1] & C, Al, N & 8.170 & 4.5 & 15.2 & -4.9 & -16.2 & 67.5 & 67.5  \\
grp/hAlN & Al, N & 8.172 & 89.1 & 9.5 & 5.4 & -5.6 & 81.0 & 81.0 \\
grp/hAlN & C & 8.172 & 87.5 & 5.1 & -10.0 & -10.0 & 0 & 0 \\
hAlN/grp/hAlN & C, Al, N & 8.077 & 10.0 & 0 & 3.2 & 3.2 & 0 & 0\\
\hline
grp/hBN ($4\times 4$) & C, B, N & 8.206 & 13.3\footnotemark[2] & 7.9 & -8.8 & -5.8 & 25.6 & 12.1 \\
grp/hBN ($4\times 4$)\footnotemark[1] & C, B, N & 8.206 & 13.4\footnotemark[2] & 7.9 & -8.8 & -5.8 & 25.9 & 12.2  \\
grp/hBN ($5\times 5$) & C, B, N & 8.246 & 11.3\footnotemark[2] & 6.8 & 8.1 & 5.4 & 23.9 & 12.7 \\
\hline
hGaN/grp/hBN & C, B, Ga, N & 8.301 & 12.0\footnotemark[2] & -69.7 & -9.7 & -6.7 & -123.3 & -116.1 \\
hAlN/grp/hBN & C, B, Al, N & 8.151 & 10.8\footnotemark[2] & 7.0 & -2.4 & -10.9 & 3.6 & 3.6\\
\end{tabular}
\end{ruledtabular}
\footnotetext[1]{The rippled structure from the relaxation of internal atomic positions.}
\footnotetext[2]{For the graphene/hBN structures, the sublattice-symmetry breaking is much larger (see Ref.\ \citenum{Zollner2019:PRB}) and the parameter $\Delta$ is given in meV.}
\end{table*}
%-----------------------------------------------------------------

The fit parameters are summarized in Table~\ref{tab:fit}.
We find that hGaN induces larger band splittings in graphene than hAlN, as reflected in the four-times larger Rashba parameter. The origin of this sizeable Rashba coupling is the deformation of C $p_z$ orbitals along the $z$ direction \cite{Gmitra2009:PRB} (see Appendix~\ref{app:B}), being much more pronounced in the hGaN case compared to the other substrates. 
In Table~\ref{tab:fit} we also summarize fit results for the graphene/hBN stacks ($4\times 4$ and $5\times 5$ supercells). Both give similar results, as already found in Ref.\ \citenum{Zollner2019:PRB}, which further validates our DFT calculations. 
In contrast to the graphene/hAlN and graphene/hGaN stacks, the staggered potential gap parameter $\Delta$ is much larger in the graphene/hBN case.
In the case of hBN, the individual graphene sublattices (A and B) are perfectly aligned above the boron and hollow positions of hBN, see Fig.~\ref{Fig:AlN_GaN_Structures}(a), leading to a strong sublattice-symmetry breaking and a large staggered potential gap. 
In contrast, for hGaN and hAlN the sublattice atoms of graphene are nearly arbitrarily aligned above the substrates, see Fig.~\ref{Fig:AlN_GaN_Structures}(b,c), and very little sublattice asymmetry arises.

We additionally calculated the low energy band structures when SOC is artificially turned off in the nitrogen, gallium, aluminum or carbon atoms.
The fit parameters for these situations are also given in Table~\ref{tab:fit}. 
From these fits we can resolve the atomic contributions to the SOC parameters induced solely by the substrate. 
When SOC of the substrate is neglected, we recover the intrinsic SOC of pristine graphene \cite{Gmitra2009:PRB}, and strongly reduce the Rashba and PIA contributions to the band splittings.

%------------------------------------------------------------
\subsection{Distance Study}
%------------------------------------------------------------

%------------------------------------------------------------------------
\begin{figure}[htb]
 \includegraphics[width=.99\columnwidth]{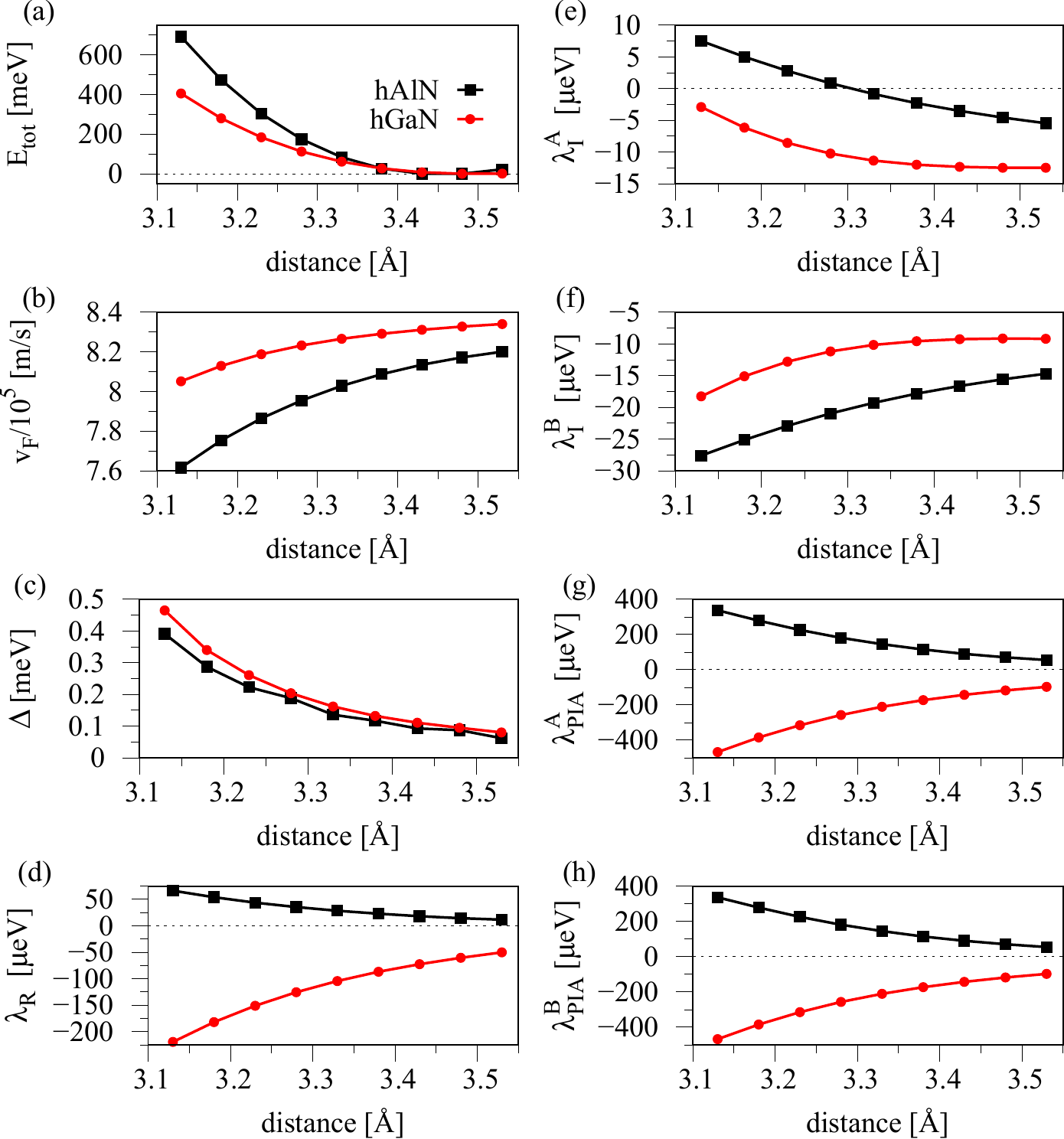}
 \caption{(Color online) Fit parameters as a function of distance between graphene and hAlN or hGaN. 
 (a) Total energy, (b) Fermi velocity $v_{\textrm{F}}$ (c) gap parameter $\Delta$, (d) Rashba SOC parameter $\lambda_{\textrm{R}}$, (e) intrinsic SOC parameter $\lambda_{\textrm{I}}^\textrm{A}$ for sublattice A, (f) intrinsic SOC parameter $\lambda_{\textrm{I}}^\textrm{B}$ for sublattice B, (g) PIA SOC parameter $\lambda_{\textrm{PIA}}^\textrm{A}$ for sublattice A, and (h) PIA SOC parameter $\lambda_{\textrm{PIA}}^\textrm{B}$ for sublattice B.
 }\label{Fig:distances}
\end{figure}
%------------------------------------------------------------------------

In Fig.~\ref{Fig:distances}, we show the evolution of the fit 
parameters for graphene/hAlN and graphene/hGaN as we vary the interlayer distance, similar to Ref.\ \cite{Zollner2019:PRB} for the graphene/hBN stack.
We find that the total energy of the systems is minimized for interlayer distances of 3.48~\AA, in agreement with the literature showing weak vdW bonding \cite{Sun2017:APL}. 
Additionally, we find that the Rashba and PIA parameters decrease with increasing distance, while the intrinsic SOC parameters converge towards the known value of about $10~\upmu$eV for pristine graphene \cite{Gmitra2009:PRB}.
The different Fermi velocities for the two systems can be attributed to the different graphene lattice constants used in the heterostructure calculations, which affects the magnitude of the nearest-neighbor hopping integral.

%------------------------------------------------------------
\subsection{Encapsulated Structures}
%------------------------------------------------------------

In the symmetric encapsulated graphene heterostructures, namely hAlN/graphene/hAlN and hGaN/graphene/hGaN, we find that, compared to the non-encapsulated cases, Rashba SOC is strongly suppressed because inversion symmetry is nearly restored, as shown in Table~\ref{tab:fit}. 
Thus, the encapsulated geometries should in principle lead to larger spin lifetimes, as is the case for graphene/hBN structures \cite{Zollner2019:PRB}.
In the case of the hBN/graphene/hGaN and hBN/graphene/hAlN structures, the fit parameters are also summarized in Table~\ref{tab:fit} and result from an interplay between the different top and bottom encapsulation layers. 
The hBN layer significantly enhances the orbital gap parameter $\Delta$ due to commensurability with the graphene lattice and the resulting sublattice-symmetry breaking. 
In the case of hBN/graphene/hGaN, the SOC parameters are similar to the non-encapsulated graphene/hGaN case, as if the hBN layer would have no effect. In particular, the Rashba and PIA SOC parameters have roughly the same values.
In contrast, for the hBN/graphene/hAlN structure, the formerly large PIA parameters from the non-encapsulated graphene/hAlN case are now strongly suppressed, while the intrinsic and Rashba SOC are of similar magnitude.

%------------------------------------------------------------
\subsection{Electric Field Effects}
%------------------------------------------------------------

%------------------------------------------------------------------------
\begin{figure}[htb]
 \includegraphics[width=.99\columnwidth]{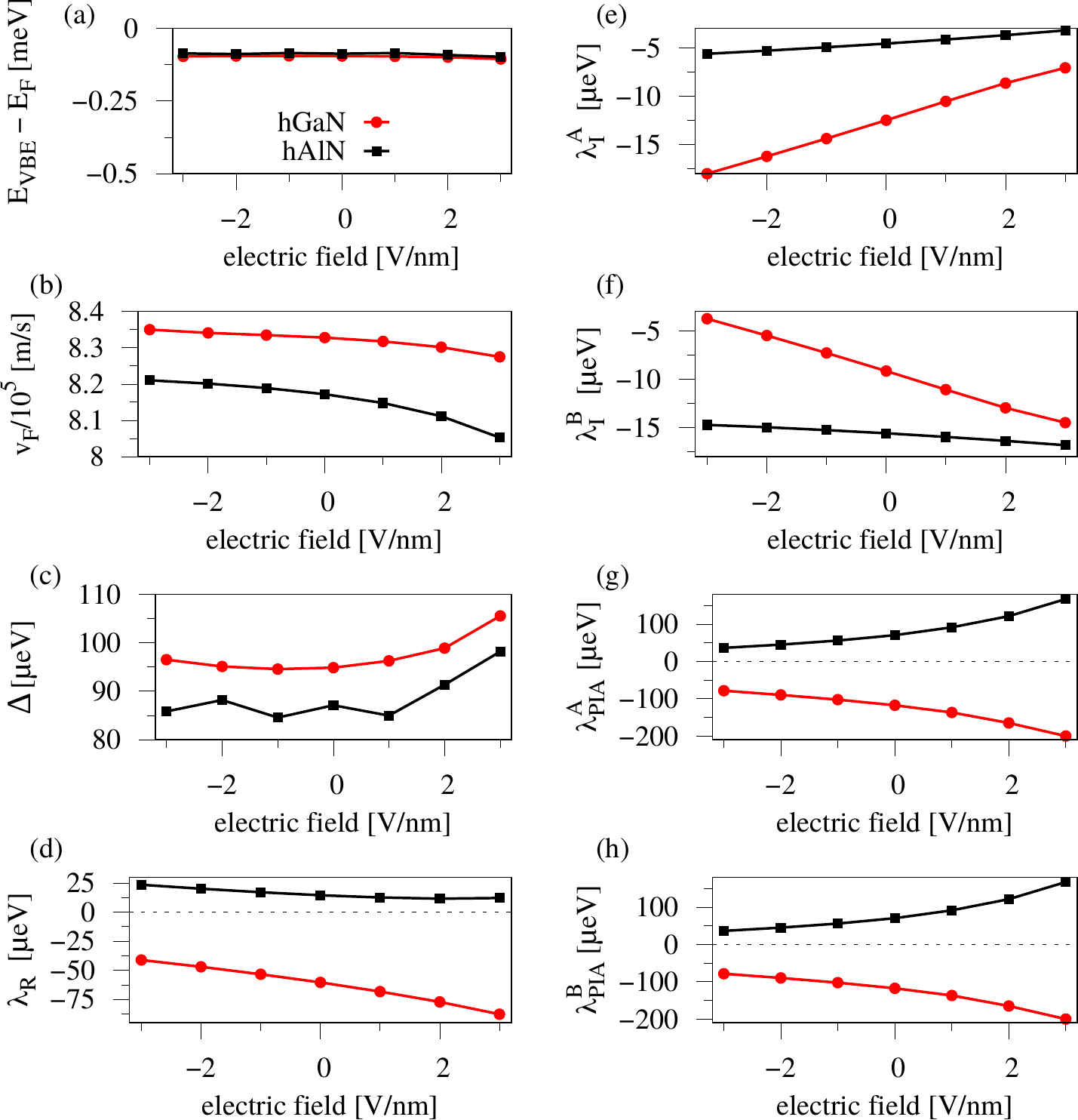}
 \caption{(Color online) Fit parameters as a function of the transverse electric field for graphene/hAlN and graphene/hGaN at fixed interlayer distances of 3.48~\AA.
 (a)~The valence band edge with respect to the Fermi level, (b)~the Fermi velocity $v_{\textrm{F}}$, (c) gap parameter $\Delta$, (d) Rashba SOC parameter $\lambda_{\textrm{R}}$, (e) intrinsic SOC parameter $\lambda_{\textrm{I}}^\textrm{A}$ for sublattice A, (f) intrinsic SOC parameter $\lambda_{\textrm{I}}^\textrm{B}$ for sublattice B, (g)~PIA SOC parameter $\lambda_{\textrm{PIA}}^\textrm{A}$ for sublattice A, and (h) PIA SOC parameter $\lambda_{\textrm{PIA}}^\textrm{B}$ for sublattice B.
 }\label{Fig:Electric_Field}
\end{figure}
%------------------------------------------------------------------------

Gating can be used to tune the SOC, especially the Rashba SOC, in the Dirac bands of graphene \cite{Gmitra2009:PRB,Zollner2019:PRB}. 
Tuning the SOC can have a significant impact on spin transport and relaxation properties. Consequently, gating is a potentially efficient control knob in experimental studies.
Here we apply a transverse electric field, modeled by a sawtooth potential, across graphene/hAlN and graphene/hGaN heterostructures for a fixed interlayer distance of 3.48~\AA. The results are summarized in Fig.~\ref{Fig:Electric_Field}. 

Indeed, most of the parameters can be tuned by the field. 
Especially in the case of hGaN, the tunability of the Rashba, intrinsic, and PIA SOC parameters is giant within our considered field limits. 
Tuning the field from $-3$ to $3$~V/nm, we find that the Rashba SOC can be enhanced in magnitude from about $40$ to $90~\upmu$eV. The intrinsic SOC parameter for sublattice A (B) decreases (increases) in magnitude, while both PIA parameters can be drastically enhanced by a factor of 2 when the field is tuned from negative to positive values.
In the case of hAlN as the substrate, the electric field tunability is somewhat similar, but not as pronounced as in the case of hGaN. However, the PIA couplings show also a strong tunability ranging from $40$ to $170~\upmu$eV within our electric field limits.

%------------------------------------------------------------
\subsection{Rippling Effects}
%------------------------------------------------------------

So far, we only considered flat monolayers stacked on top of each other, with fixed interlayer distances. 
In experiment, the layers are typically not completely flat and rippling occurs due to imperfections and impurities. 
How does rippling affect the proximity-induced SOC?
The starting point are the graphene/hXN heterostructures with the energetically most favorable interlayer distance of 3.48~\AA, to which we apply the force relaxation \cite{Yu1991:PRB,Kohler1996:CPC,Madsen2001:PRB} implemented in the WIEN2k code. Thereby, we determine the equilibrium positions of all atoms by minimizing the forces on the nuclei, while respecting the $C_{3v}$-symmetry of the heterostructures. 
When all forces are below 0.5~mRy/bohr, we continue to calculate the low energy bands and apply the model Hamiltonian fit routine. The fit results are summarized in Table~\ref{tab:fit}. 

After relaxation of a given graphene/hXN heterostructure, we also calculate the mean values, $\Bar{z}$, and the standard deviations, $\Delta z$, of the $z$ coordinates of the atoms belonging to the individual monolayers. 
From the mean values, we calculate the average interlayer distance, \mbox{$d = \Bar{z}_{\textrm{grp}}-\Bar{z}_{\textrm{hXN}}$}, between graphene and hXN, while the standard deviations represent the amount of rippling of the layers.
The results are summarized in Table~\ref{tab:rippling}. 
Overall, we find that the average interlayer distances are barely affected and the individual monolayers are still nearly flat with a maximum rippling below 1~pm. 
Comparing the fit parameters in Table~\ref{tab:fit}, for the flat and rippled structures, we find that mainly the staggered potential gap $\Delta$ is affected, while the SOC parameters stay nearly the same. In the case of hGaN, $\Delta$ is about twice as large in the flat structure compared to the rippled one. 
For hAlN, the effect is even more drastic, since the rippled structure nearly leads to a vanishing sublattice asymmetry in graphene, while the flat structure has a 20-times larger asymmetry, as reflected in the parameter $\Delta$. 

%-----------------------------------------------------------------
\begin{table}[htb]
\caption{\label{tab:rippling} The calculated average interlayer distances, $d$, and the ripplings, $\Delta z$, of the monolayers, of the graphene/hXN heterostructures after relaxation of internal forces.  }
\begin{ruledtabular}
\begin{tabular}{lccc}
System & $d$ [\AA] &  $\Delta z_{\textrm{grp}}$ [pm] &  $\Delta z_{\textrm{hXN}}$ [pm]\\
\hline
grp/hBN  & 3.47952  & 0.005 & 0.117  \\
grp/hAlN & 3.48017  & 0.089 & 0.282 \\
grp/hGaN & 3.47997  & 0.042 & 0.093  \\
\end{tabular}
\end{ruledtabular}
\end{table}
%-----------------------------------------------------------------

%------------------------------------------------------------
\section{Spin Relaxation}
%------------------------------------------------------------

\subsection{Model and Numerical Approach}

Spin relaxation in graphene is a vibrant topic in condensed matter physics \cite{Leutenantsmeyer2018:PRL, Xu2018:PRL,Zihlmann2018:PRB, Omar2018:PRB,Ertler2009:PRB,Huertas2009:PRL,Zhang2012:NJP,Cummings2017:PRL,Offidani2018:arxiv,Raes2016:NC, Ringer2018:PRB, Zhu2018:PRB,Irmer2018:PRB, Kochan2014:PRL, Miranda2017:JPCS,Zihlmann2018:PRB, Omar2018:PRB,Han2011:PRL,Han2012:NL,Yang2011:PRL,Tombros2007:Nat}.
Therefore, we now use a combination of modeling and numerical simulations to predict the nature and magnitude of spin relaxation in graphene/hXN heterostructures. 
To numerically simulate spin transport, we use a real-space time-dependent approach that has been employed for the study of electrical and spin transport in a wide variety of disordered materials \cite{Fan2019:arxiv}. We initialize a spin-polarized random-phase state in a 500$\times$500 nm$^2$ sample of graphene, and we evolve it in time using an efficient Chebyshev expansion of the time evolution operator. At each time step we calculate the spin polarization $s_i(E,t)$ and the mean-square displacement $\Delta X^2(E,t)$ of the state as a function of the Fermi energy $E$. The energy-dependent spin lifetime $\tau_{\text{s},i}(E)$ is then extracted by fitting the spin polarization to an exponential decay, $s_i(E,t) = \exp(-t/\tau_{\text{s},i}(E))$. We also calculate the momentum relaxation time $\tau_\text{p}(E) = 2D(E)/v_\text{F}^2$, where $D(E) = \max\limits_t \left\{ \frac{1}{2}\frac{\text{d}}{\text{d}t} \Delta X^2(E,t) \right\}$ is the semiclassical diffusion coefficient.

To induce charge scattering and thus spin relaxation in our simulations, we include a random distribution of Gaussian-shaped electrostatic impurities which are meant to represent the impact of charged impurity scattering in graphene \cite{Chen2008:NatPhys, Adam2009:PRB}. The electrostatic potential at each atomic site $i$ is then given by $\epsilon_i = \sum\limits_j V_j \exp(-|\vec{r}_i-\vec{r}_j|^2 / 2\xi^2)$, where $\vec{r}_i$ is the position of each carbon atom, $\vec{r}_j$ is the position of each impurity, $\xi$ is the width of each impurity, and the height $V_j$ of each impurity is randomly distributed in $[-V,V]$. Here we use $V = 2.8$ eV, $\xi = \sqrt{3}a$, and an impurity density of $0.1\%$. This choice of parameters leads to a charge mobility around $\mu \approx 1000$ cm$^2$/Vs and intervalley scattering in the range of $\tau_\text{iv} \approx (10-60)\tau_\text{p}$. Here we should note that this mobility is quite a bit lower than what would be expected for graphene interfaced with these hXN materials. However, limitations on the sample size we can simulate require us to use this value to ensure proper numerical convergence. Assuming the Dyakonov-Perel mechanism of spin relaxation (see below), the obtained spin lifetimes can be scaled to expected experimental mobilities according to $\tau_\text{s}^{-1} \propto \mu$. Finally, for each graphene/hXN heterostructure, we convert the continuum Hamiltonian of Eq.\ \eqref{Eq:Hamiltonian} to a real-space tight-binding form and use the first set of parameters in Table \ref{tab:fit}.

In our numerical simulations we have only considered the graphene/hGaN heterostructure, as the others have lower SOC, putting their spin lifetimes outside the range of what we can simulate numerically. Thus, to explain the features seen in our simulations and to make predictions for the hAlN and hBN heterostructures, we now describe a simple and transparent model of spin relaxation in these systems. We start by assuming the only source of spin relaxation is from the SOC parameters defined in Eq.\ \eqref{Eq:Hamiltonian}, and not from extrinsic sources such as ripples or magnetic impurities \cite{Huertas2009:PRL, Kochan2014:PRL}. There are two traditional spin relaxation mechanisms driven by uniform SOC, namely the Elliott-Yafet (EY) and the Dyakonov-Perel (DP) mechanisms \cite{Elliott1954:PR, Yafet1963:SSP, Dyakonov1972:SPSS}. In the EY mechanism, spin flips occurs at the point of scattering, and the spin lifetime is proportional to the scattering time. In graphene, the EY mechanism is predicted to relax in-plane spins via the intrinsic SOC \cite{Ochoa2012:PRL}. In the DP mechanism, SOC-induced spin precession and dephasing occur between scattering events, leading to an inverse scaling between spin lifetime and scattering time.

The numerical simulations show a negligible EY contribution, primarily arising from the small value of $\lambda_\text{I}$ in all systems. Thus, we focus solely on the DP mechanism. The spin relaxation rates in graphene are then given by \cite{Cummings2017:PRL, Garcia2018:CSR, Zollner2019:PRB}
\begin{align}
\label{eq:relaxation_rate}
\tau_{\text{s},x}^{-1} &= \tau_\text{p} \langle \omega_y^2 \rangle + \tau_\text{iv} \langle \omega_z^2 \rangle, \nonumber \\
\tau_{\text{s},y}^{-1} &= \tau_\text{p} \langle \omega_x^2 \rangle + \tau_\text{iv} \langle \omega_z^2 \rangle, \\
\tau_{\text{s},z}^{-1} &= \tau_\text{p} \langle \omega_x^2 + \omega_y^2 \rangle, \nonumber
\end{align}
where $\tau_{\text{s},x/y}$ are the lifetimes of spins pointing in the graphene plane, $\tau_{\text{s},z}$ is the out-of-plane spin lifetime, $\tau_\text{p}$ is the momentum relaxation time, $\tau_\text{iv}$ is the intervalley scattering time, and $\omega_i$ are the components of the effective magnetic field induced by SOC. As we will see below, the $z$-component of the effective magnetic field has opposite sign in the $\pm$K valleys and thus drives spin relaxation through intervalley scattering, while the in-plane components depend on the direction of electron momentum and thus are connected to $\tau_\text{p}$.

The effective magnetic field can be written as (see Appendix ~\ref{app:A})
\begin{align}
\label{eq:Bsoc}
\frac{\hbar}{2}\omega_x &= \left[ \left(\pm \lambda_\text{R} + ak\lambda_\text{PIA}^+ \right) + g_\parallel(k) \right] \cdot \sin(-\theta), \nonumber \\
\frac{\hbar}{2}\omega_y &= \left[ \left(\pm \lambda_\text{R} + ak\lambda_\text{PIA}^+ \right) + g_\parallel(k) \right] \cdot \cos(\theta), \nonumber \\
\frac{\hbar}{2}\omega_z &= \left[ \lambda_\text{VZ} + g_z(k) \right] \cdot \tau,
\end{align}
where $\lambda_\text{PIA}^+ = (\lambda_\text{PIA}^\text{A} + \lambda_\text{PIA}^\text{B})/2$ gives rise to a $k$-linear spin splitting similar to Rashba SOC in traditional 2D electron gases, $\lambda_\text{VZ} = (\lambda_\text{I}^\text{A} - \lambda_\text{I}^\text{B})/2$ is the valley-Zeeman SOC, and $\theta$ denotes the direction of electron momentum $k$ in the graphene plane. The $+(-)$ sign in $\omega_{x/y}$ is for the valence (conduction) band, and thus a strong PIA SOC can induce electron-hole asymmetry in the spin relaxation. Note that $\omega_z$ is proportional to $\tau$ and thus has opposite sign in opposite valleys, as mentioned above. The second terms in the brackets, $g_{\parallel / z}(k)$, represent higher-order terms in the effective magnetic field. These are large near the graphene Dirac point ($\hbar v_\text{F} k \lesssim \Delta, \lambda_i$) and decay as $\sim 1/k$ at higher energies.

\subsection{Single-Sided Heterostructures}

\begin{figure}[htb]
 \includegraphics[width=0.99\columnwidth]{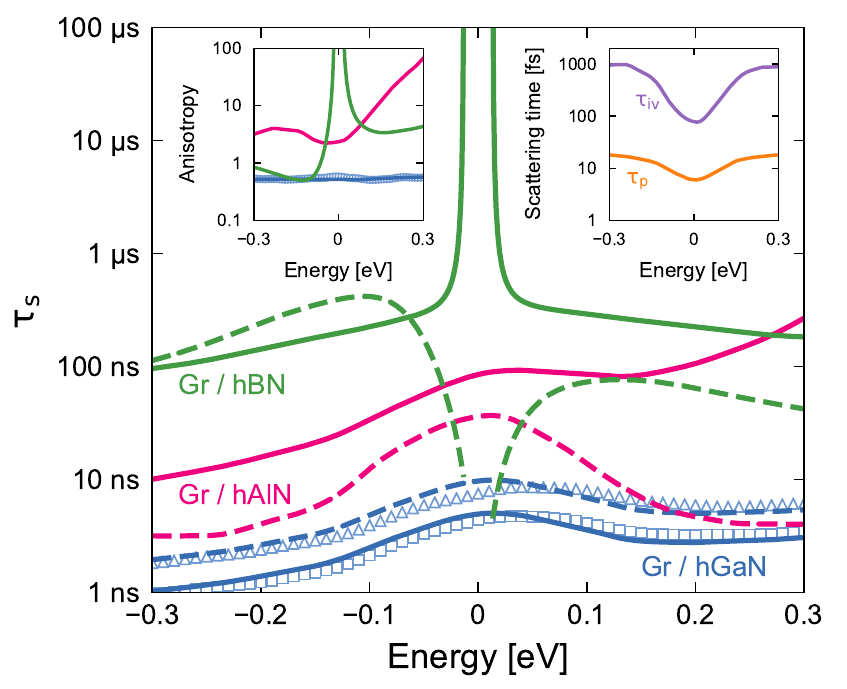}
 \caption{(Color online) Predicted spin lifetime in graphene/hXN heterostructures. The solid (dashed) lines show the out-of-plane (in-plane) spin lifetimes calculated from Eqs.\ \eqref{eq:relaxation_rate}-\eqref{eq:terms_BN}. The symbols show the spin lifetime from numerical simulations of graphene/hGaN. The left inset shows the spin lifetime anisotropy for each heterostructure, and the right inset shows the momentum relaxation time and intervalley scattering time calculated from numerical simulations.
 }\label{fig:DP_pred}
\end{figure}

The numerical results for the graphene/hGaN heterostructure are shown as the symbols in Fig.\ \ref{fig:DP_pred}. The right inset shows the momentum relaxation time, which ranges from 6 to 20 fs, and the intervalley scattering time, which ranges from 75 fs to 1 ps. The blue triangles (squares) in the main panel show the in-plane (out-of-plane) spin lifetime as a function of Fermi energy. Lifetimes are between 1 and 10 ns, with a maximum around the charge neutrality point and a slight electron-hole asymmetry. The spin lifetime anisotropy, defined as the ratio of out-of-plane to in-plane spin lifetime, $\zeta \equiv \tau_\text{s,z} / \tau_\text{s,x}$, is shown as the blue circles in the left inset. The anisotropy is nearly $1/2$ over the entire energy range, indicating that spin relaxation is driven by the DP mechanism with a dominant Rashba+PIA SOC. The peak in the spin lifetime around $E=0$, corresponding to the dip in $\tau_\text{p}$ at the same point, is also indicative of the DP mechanism.

%-----------------------------------------------------------------------------
%-----------------------------------------------------------------------------

To better understand these results, we now turn to the model of Eqs.\ \eqref{eq:relaxation_rate}-\eqref{eq:Bsoc}. We first examine the cases of graphene/hAlN and graphene/hGaN, for which all energies in Table \ref{tab:fit} are small, less than 100 $\upmu$eV (for PIA SOC, the relevant energy scale is $ak\lambda_\text{PIA}$). If we consider spin transport near room temperature ($k_\text{B}T \approx 26$ meV), then the $g_{\parallel / z}(k)$ terms in Eq.\ \eqref{eq:Bsoc} are smeared out by the large thermal broadening and only the first terms in the brackets are responsible for spin transport. The in-plane and out-of-plane spin relaxation rates are then given by
\begin{align}
\label{eq:lifetime_GaAlN}
\tau_{\text{s},x}^{-1} &= \tau_{\text{s},y}^{-1} = \frac{1}{2} \tau_{\text{s},z}^{-1} + \left( \frac{2}{\hbar} \lambda_\text{VZ} \right)^2 \tau_\text{iv}, \nonumber \\
\tau_{\text{s},z}^{-1} &= \left(\frac{2}{\hbar} \left( \pm \lambda_\text{R} + ak\lambda_\text{PIA}^+ \right) \right)^2 \tau_\text{p},
\end{align}
the same as what was previously derived for graphene/TMDC heterostructures, in which strong valley-Zeeman SOC coupled with intervalley scattering leads to large values of spin lifetime anisotropy \cite{Cummings2017:PRL, Garcia2018:CSR, Offidani2018:arxiv}. Plugging the parameters from Table \ref{tab:fit} and the values of $\tau_\text{p}$ and $\tau_\text{iv}$ from the numerical simulations into Eq.\ \eqref{eq:lifetime_GaAlN}, we obtain spin lifetimes for graphene/GaN that match very well with the numerical simulations, as shown by the blue lines in Fig.\ \ref{fig:DP_pred}. The spin lifetime anisotropy of $\zeta \approx 1/2$ also matches very well, confirming the dominance of Rashba+PIA over valley-Zeeman SOC.

The red lines in Fig.\ \ref{fig:DP_pred} show the expected spin lifetimes for graphene/AlN. Here the Rashba SOC is somewhat smaller, allowing the PIA SOC to play a more dominant role and resulting in significant electron-hole asymmetry. In particular, for positive energies the PIA cancels out the Rashba SOC, which enhances the out-of-plane spin lifetime. Meanwhile, the in-plane spin lifetime is suppressed by an enhanced contribution from valley-Zeeman SOC, leading to large spin lifetime anisotropy. For negative energies, PIA augments the Rashba SOC and the anisotropy is reduced.

Now we consider the case of spin relaxation in graphene/hBN. In this system the staggered sublattice potential $\Delta$ is large, opening a band gap on the order of the thermal broadening, and thus the higher-order terms in Eq.\ \eqref{eq:Bsoc} cannot be ignored. These terms are complicated expressions of the various parameters in the Hamiltonian (see Appendix~\ref{app:A}), but for the values given in Table \ref{tab:fit} they are well captured (to within 1\%) by
\begin{align}
\label{eq:terms_BN}
g_\parallel(k) &= \frac{\Delta^2\lambda_\text{VB} + 2\Delta\varepsilon_\text{PIA}^-\varepsilon_k^\Delta}{\left(\varepsilon_k^\Delta\right)^2}, \nonumber \\
g_z(k) &= \frac{\Delta^2\lambda_\text{VZ} + 2\Delta\lambda_\text{I}\varepsilon_k^\Delta}{\left(\varepsilon_k^\Delta\right)^2},
\end{align}
where $\lambda_\text{VB} = \lambda_\text{R} + ak\lambda_\text{PIA}^+$, $\varepsilon_\text{PIA}^- = ak(\lambda_\text{PIA}^\text{A}-\lambda_\text{PIA}^\text{B})/2$, $\lambda_\text{I} = (\lambda_\text{I}^\text{A}+\lambda_\text{I}^\text{B})/2$ is the intrinsic or Kane-Mele SOC, and $\varepsilon_k^\Delta = \hbar v_\text{F} k + \sqrt{(\hbar v_\text{F}k)^2 + \Delta^2}$. Equation \eqref{eq:terms_BN} was derived for spin transport in the conduction band, but similar terms apply to the valence band by replacing $\lambda_\text{R} \rightarrow -\lambda_\text{R}$ and $\varepsilon_k^\Delta \rightarrow -\varepsilon_k^\Delta$.

In the high-energy limit, the terms in Eq.\ \eqref{eq:terms_BN} disappear and the spin lifetime is described by Eq.\ \eqref{eq:lifetime_GaAlN}. However, as the Fermi energy approaches the band edges ($k \rightarrow 0$), $\omega_{x/y} \rightarrow 0$ and $\omega_z \rightarrow 2(\lambda_\text{I}+\lambda_\text{VZ})$. Thus, the out-of-plane spin lifetime diverges while the in-plane spin lifetime remains finite, leading to a giant spin lifetime anisotropy around the charge neutrality point. This behavior is shown as the green lines in Fig.\ \ref{fig:DP_pred}, and corroborates previous predictions of spin lifetime anisotropy in these systems \cite{Zollner2019:PRB}. These results indicate that a large spin lifetime anisotropy should be visible around the charge neutrality point in graphene heterostructures with a large staggered sublattice potential, even if the SOC is small.

\subsection{Electric Field Dependence}

\begin{figure}[htb]
 \includegraphics[width=0.99\columnwidth]{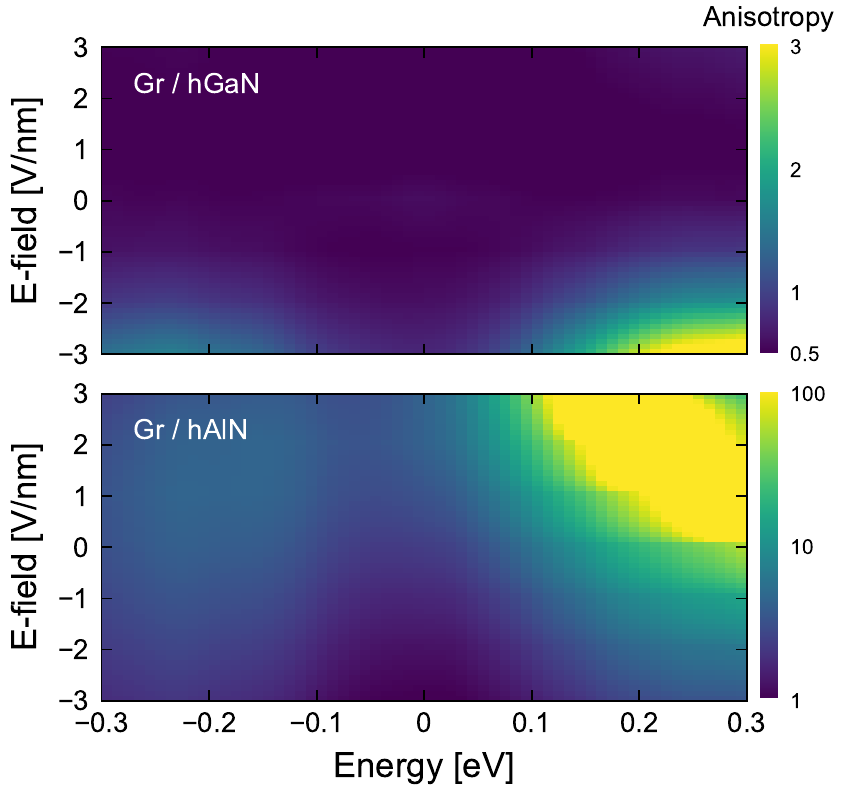}
 \caption{(Color online) Spin lifetime anisotropy in graphene/hGaN and graphene/hAlN heterostructures. Results for graphene/hGaN were calculated numerically, while those for graphene/hAlN were derived from Eq.\ \eqref{eq:lifetime_GaAlN}.}
 \label{fig:ts_E_field}
\end{figure}

As shown in Fig.\ \ref{Fig:Electric_Field} above, an external electric field can be used to efficiently tune the various spin-orbit parameters in the graphene/hGaN and graphene/hAlN heterostructures. We plot the resulting impact on the spin lifetime anisotropy in Fig.\ \ref{fig:ts_E_field}. For positive E-fields, the anisotropy in graphene/hGaN remains around 1/2, as the Rashba SOC is the dominant term. For negative fields, the Rashba SOC strength is reduced, valley-Zeeman SOC begins to play a role, and a modest anisotropy appears, reaching a value of $\zeta \approx 3$ at high doping levels.

In contrast, the Rashba SOC in graphene/hAlN is somewhat weaker and has the opposite sign. The result is that the spin lifetime anisotropy is significantly larger (note the log color scale for this system), and is maximized at positive E-fields, where the magnitude of Rashba SOC is smallest. Although the magnitude of the anisotropy is quite different, it is evident that in both systems it is tunable via external field and doping.

\subsection{Double-Sided Heterostructures}

Beyond an external electric field, spin transport properties may be further tuned by sandwiching graphene between different hXN layers. For example, by placing hBN on one side of graphene and hGaN or hAlN on the other side, one can realize a system that combines the SOC properties of graphene/hGaN or graphene/hAlN, but with the large orbital gap of graphene/hBN. This can be seen by examining the final two rows of Table \ref{tab:fit}. 

\begin{figure}[htb]
 \includegraphics[width=0.99\columnwidth]{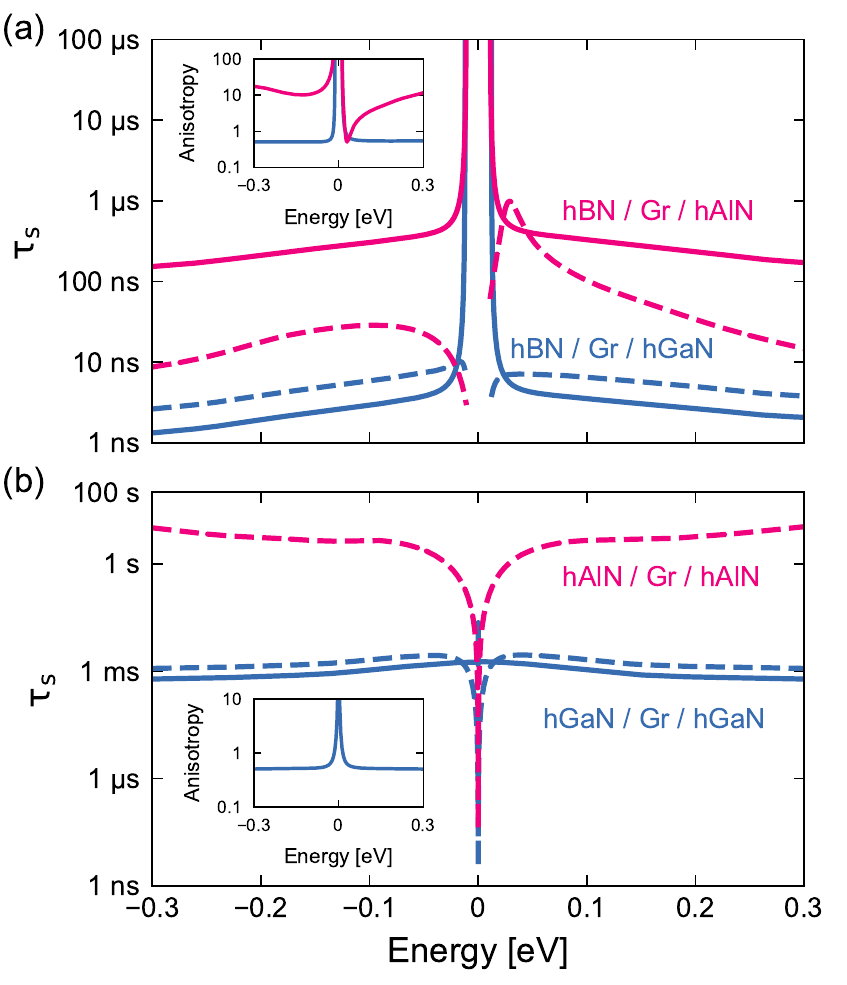}
 \caption{(Color online) Predicted spin lifetime in (a) hBN/graphene/hXN and (b) hXN/graphene/hXN heterostructures. The solid (dashed) lines show the out-of-plane (in-plane) spin lifetimes calculated from Eqs.\ \eqref{eq:relaxation_rate}-\eqref{eq:Bsoc_encaps}. The insets show the corresponding spin lifetime anisotropy.}
 \label{fig:DP_pred_double}
\end{figure}

Spin lifetimes for the hBN/graphene/hGaN and hBN/graphene/hAlN systems are shown in Fig.\ \ref{fig:DP_pred_double}(a), with solid (dashed) lines denoting out-of-plane (in-plane) lifetime. The spin lifetime anisotropy is shown in the inset. As expected, each system exhibits a giant anisotropy around the charge neutrality point, due to the orbital gap induced by the hBN layer. Away from the gap, the hBN/graphene/hGaN system behaves just like the single-sided graphene/hGaN system, with an anisotropy of 1/2 resulting from the dominant Rashba SOC. Meanwhile, the hBN/graphene/hAlN system exhibits large anisotropy and strong electron-hole asymmetry, similar to Fig.\ \ref{fig:DP_pred} above. Thus, these double-sided heterostructures appear to behave as a ``superposition'' of two single-sided heterostructures, with hBN providing the orbital gap and hGaN or hAlN providing the particular nature of the SOC.

In Figs.\ \ref{fig:DP_pred} and \ref{fig:DP_pred_double}(a) we have predicted spin lifetimes ranging from 1~ns up to hundreds of ns. These large lifetimes arise from the small SOC in these systems, but as shown in Table \ref{tab:fit}, the SOC can be further reduced by encapsulating graphene between two identical hXN layers. In this case, the valley-Zeeman and PIA SOC disappear, and the Rashba SOC is also completely or nearly eliminated. In this situation the effective magnetic field becomes
\begin{align}
\label{eq:Bsoc_encaps}
\frac{\hbar}{2}\omega_x &= \pm \lambda_\text{R} \left[ 1 + 2\lambda_\text{I}/\varepsilon_k^{\Delta\text{I}} \right] \cdot \sin(-\theta), \nonumber \\
\frac{\hbar}{2}\omega_y &= \pm \lambda_\text{R} \left[ 1 + 2\lambda_\text{I}/\varepsilon_k^{\Delta\text{I}} \right]  \cdot \cos(\theta), \nonumber \\
\frac{\hbar}{2}\omega_z &= \Delta\left[ 2\lambda_\text{I}/\varepsilon_k^{\Delta\text{I}} \right] \cdot \tau,
\end{align}
where $\varepsilon_k^{\Delta\text{I}} = \hbar v_\text{F} k + \sqrt{(\hbar v_\text{F}k)^2 + (\Delta + \lambda_\text{I})^2}$. In Fig.\ \ref{fig:DP_pred_double}(b) we plot the spin lifetimes for graphene encapsulated by hGaN or hAlN. Owing to the small SOC, all spin lifetimes are extremely large; around one millisecond for GaN and up to several seconds for AlN. In these systems, $\omega_z$ strongly increases at low energies, leading to a sharp decrease of the in-plane spin lifetime and thus a large anisotropy around the charge neutrality point for the GaN system. For the AlN system, the out-of-plane spin lifetime is infinite due to the absence of Rashba SOC and thus is not pictured.

%-----------------------------------------------------------------------------
%-----------------------------------------------------------------------------

It should be noted that here we have used the same disorder as before, but in these encapsulated systems $\tau_\text{p}$ is expected to be up to 100 times higher, corresponding to a mobility around 100,000 cm$^2$/Vs. This would reduce the spin lifetimes by a factor of 100. 

Finally, we would also like to comment on the extremely large values of spin lifetime reported in Fig.\ \ref{fig:DP_pred_double}, as well as the diverging lifetimes and anisotropies seen in Figs.\ \ref{fig:DP_pred} and \ref{fig:ts_E_field}. These values are reached because we are only considering the DP mechanism of spin relaxation, which leads to large spin lifetimes when SOC becomes small. However, such values will never be reached, as other fundamental sources of spin relaxation -- such as corrugations or phonons -- are expected to take over once DP relaxation becomes sufficiently weak. Therefore, hXN/graphene/hBN or symmetric hXN/graphene/hXN heterostructures may be suitable systems for exploring these other spin relaxation mechanisms, and their role as an ultimate upper bound on spin transport in graphene.

%------------------------------------------------------------
\section{Summary}
%------------------------------------------------------------

In this work, we have considered heterostructures of graphene and the monolayer hexagonal nitride insulators, hBN, hAlN, and hGaN, from first principles. 
In the cases of graphene on hAlN and hGaN, we performed an interlayer distance study to find the energetically most favorable van der Waals gap, which is about 3.48~\AA. 
Our calculations, combined with a low energy model Hamiltonian, reveal that graphene's linear dispersion is well preserved within the insulator band gap for all considered heterostructures, and the fitted model SOC parameters, describing the spin splitting of the Dirac states, are below 100~$\upmu$eV for optimized interlayer distances.
From the interlayer distance study, we also find that the fitted orbital and SOC parameters strongly depend on the distance between graphene and the substrate. 
Overall, for the non-encapsulated structures, Rashba and PIA contributions dominate the spin splitting, while they are suppressed in the encapsulated cases where intrinsic SOC terms of maximum 10~$\upmu$eV dominate.  
The model parameters are then used to estimate spin lifetimes in the encapsulated and non-encapsulated heterostructure systems. 

With respect to spin transport and relaxation, several interesting and surprising features appear in these graphene/hXN systems. Depending on the relative strength of the parameters in the Hamiltonian, spin transport can be quite conventional, can exhibit strong electron/hole asymmetry in the lifetimes and in the spin lifetime anisotropy, or can exhibit giant anisotropy only around the charge neutrality point. This enhanced anisotropy around the charge neutrality point arises from a strong staggered sublattice potential $\Delta$ in combination with another source of SOC. This suggests that the graphene/hBN system, in which sublattice symmetry breaking can open a significant band gap, is the best candidate to observe this behavior. Meanwhile, graphene encapsulated in GaN or AlN appears to be an excellent candidate for exploring the upper limits of spin transport in 2D van der Waals heterostructures \cite{Ertler2009:PRB, Huertas2009:PRL, Zhang2012:NJP}.

\acknowledgments
This work was funded by the Deutsche Forschungsgemeinschaft (DFG, German Research Foundation) SFB 1277 (Project-ID 314695032), DFG SPP 1666, DFG SPP 2244, and the European Unions Horizon 2020 research and innovation program 
under Grant No.881603 (Graphene Flagship). ICN2 is supported by the Severo Ochoa program from Spanish MINECO (grant no.\ SEV-2017-0706) and funded by the CERCA Programme / Generalitat de Catalunya.

\begin{widetext}
\appendix

\section{Derivation of Effective Magnetic Field}
\label{app:A}

In this Appendix we derive the complete form of the effective magnetic field, $\vec{\omega}$, induced by SOC. In the basis $\left\{ \ket{\Psi_{\textrm{A}}, \uparrow}, \ket{\Psi_{\textrm{A}}, \downarrow}, \ket{\Psi_{\textrm{B}}, \uparrow}, \ket{\Psi_{\textrm{B}}, \downarrow} \right\}$, the Hamiltonian of Eq.\ \eqref{Eq:Hamiltonian} can be written in matrix form as
\begin{equation}
\mathcal{H} =
\begin{bmatrix}
\Delta + \tau \lambda_\text{I}^\text{A}
&
\varepsilon_k
&
f^*\varepsilon_\text{PIA}^\text{A}
&
\varepsilon_\text{R}^+
\\
\varepsilon_k^*
&
-\Delta - \tau \lambda_\text{I}^\text{B}
&
\varepsilon_\text{R}^-
&
f^*\varepsilon_\text{PIA}^\text{B}
\\
f\varepsilon_\text{PIA}^\text{A}
&
{\varepsilon_\text{R}^-}^*
&
\Delta - \tau \lambda_\text{I}^\text{A}
&
\varepsilon_k
\\
{\varepsilon_\text{R}^+}^*
&
f\varepsilon_\text{PIA}^\text{B}
&
\varepsilon_k^*
&
-\Delta + \tau \lambda_\text{I}^\text{B}
\end{bmatrix},
\end{equation}
where $\varepsilon_k = \hbar v_\text{F} k \tau \text{e}^{\text{i}\tau\theta}$, $\varepsilon_\text{R}^\pm = \text{i}\lambda_\text{R}(\tau \pm 1)$, $\varepsilon_\text{PIA}^\text{A,B} = ak\lambda_\text{PIA}^\text{A,B}$, and $f = \text{i}\text{e}^{\text{i}\theta}$. From this we want a Hamiltonian projected onto the conduction or valence bands with the form
\begin{gather}
\mathcal{H}_\text{CB/VB} = \mathcal{H}_\text{CB/VB}^0 + \frac{1}{2}\hbar \vec{\omega} \cdot \vec{s} 
= \begin{bmatrix} \pm\varepsilon_0 & 0 \\ 0 & \pm\varepsilon_0 \end{bmatrix} + \frac{\hbar}{2} \begin{bmatrix} \omega_z & \omega_x - \text{i}\omega_y \\ \omega_x + \text{i}\omega_y & -\omega_z \end{bmatrix}.
\label{eq:ham_eff}
\end{gather}
where $\mathcal{H}_\text{CB/VB}^0$ is the graphene Hamiltonian in the absence of SOC ($\lambda = 0$) and sublattice symmetry breaking ($\Delta = 0$), $\vec{\omega}$ is the effective magnetic field induced by these terms, and $\vec{s}$ are the spin Pauli matrices.

When $\Delta = \lambda = 0$, the eigenstates of $\mathcal{H}$ can be written as
\begin{align}
\ket{\Psi_{\textrm{CB}},\uparrow} = \begin{bmatrix}1 & \tau\text{e}^{-\text{i}\tau\theta} & 0 & 0 \end{bmatrix}^\text{T} / \sqrt{2}, \nonumber \\
\ket{\Psi_{\textrm{CB}},\downarrow} = \begin{bmatrix}0 & 0 & 1 & \tau\text{e}^{-\text{i}\tau\theta} \end{bmatrix}^\text{T} / \sqrt{2}, \nonumber \\
\ket{\Psi_{\textrm{VB}},\uparrow} = \begin{bmatrix}1 & -\tau\text{e}^{-\text{i}\tau\theta} & 0 & 0 \end{bmatrix}^\text{T} / \sqrt{2}, \nonumber \\
\ket{\Psi_{\textrm{VB}},\downarrow} = \begin{bmatrix}0 & 0 & 1 & -\tau\text{e}^{-\text{i}\tau\theta} \end{bmatrix}^\text{T} / \sqrt{2},
\end{align}
where the CB/VB states have eigenenergy $\pm\varepsilon_0 = \pm\hbar v_\text{F} k$. Writing the Hamiltonian in the basis $\left\{ \ket{\Psi_\text{CB},\uparrow}, \ket{\Psi_\text{CB},\downarrow}, \ket{\Psi_\text{VB},\uparrow}, \ket{\Psi_\text{VB},\downarrow} \right\}$ yields
\begin{gather}
\label{eq:ham_cbvb}
\mathcal{H} \equiv \begin{bmatrix} \mathcal{H}_\text{CC} & \mathcal{H}_\text{CV} \\ \mathcal{H}_\text{VC} & \mathcal{H}_\text{VV} \end{bmatrix} = 
\begin{bmatrix}
\varepsilon_0 + \tau \lambda_\text{VZ}
&
f^* \lambda_\text{CB}
&
\Delta + \tau \lambda_\text{I}
&
f^* \lambda_\text{CV}^+
\\
f \lambda_\text{CB}
&
\varepsilon_0 - \tau \lambda_\text{VZ}
&
f \lambda_\text{CV}^-
&
\Delta - \tau \lambda_\text{I}
\\
\Delta + \tau \lambda_\text{I}
&
f^* \lambda_\text{CV}^-
&
-\varepsilon_0 + \tau \lambda_\text{VZ}
&
f^* \lambda_\text{VB}
\\
f \lambda_\text{CV}^+
&
\Delta - \tau \lambda_\text{I}
&
f \lambda_\text{VB}
&
-\varepsilon_0 - \tau \lambda_\text{VZ}
\end{bmatrix}, 
\end{gather}
where $\lambda_\text{CB} = -\lambda_\text{R} + ak\lambda_\text{PIA}^+$, $\lambda_\text{VB} = \lambda_\text{R} + ak\lambda_\text{PIA}^+$, and $\lambda_\text{CV}^\pm = \varepsilon_\text{PIA}^- \pm \tau\lambda_\text{R}$. In the heterostructures with hBN, $\Delta$ is much larger than the spin-orbit parameters and the eigenenergies of $\mathcal{H}$ are given by $E \approx \pm \sqrt{\varepsilon_0^2 + \Delta^2}$. When graphene is sandwiched by two identical hXN layers, Table \ref{tab:fit} indicates that VZ, Rashba, and PIA SOC disappear, and the eigenenergies of $\mathcal{H}$ are then given by $E \approx \sqrt{\varepsilon_0^2 + (\Delta \pm \lambda_\text{I})^2}$ for the conduction band and the opposite sign for the valence band. In Figs.\ \ref{fig:DP_pred} and \ref{fig:DP_pred_double} we used the eigenenergy of the upper conduction band, but the results do not change appreciably if we use the lower conduction band instead.

We now reduce the full Schr\"{o}dinger equation, $\mathcal{H}\psi = E\psi$, to expressions projected onto the conduction or valence band only,
\begin{gather}
\mathcal{H}\psi = E\psi \nonumber \\
\Downarrow  \nonumber \\
\begin{bmatrix} \mathcal{H}_\text{CC} & \mathcal{H}_\text{CV} \\ \mathcal{H}_\text{VC} & \mathcal{H}_\text{VV} \end{bmatrix} \begin{bmatrix} \psi_\text{CB} \\ \psi_\text{VB} \end{bmatrix} = E \begin{bmatrix} \psi_\text{CB} \\ \psi_\text{VB} \end{bmatrix} \nonumber \\
\Downarrow  \nonumber \\
\mathcal{H}_\text{CB}\psi_\text{CB} = E\psi_\text{CB} \nonumber \\
\mathcal{H}_\text{VB}\psi_\text{VB} = E\psi_\text{VB} \nonumber \\
\text{where}  \nonumber \\
\mathcal{H}_\text{CB} \equiv \left(\mathcal{H}_\text{CC}+\mathcal{H}_\text{CV}\left(E-\mathcal{H}_\text{VV}\right)^{-1} \mathcal{H}_\text{VC}\right) \nonumber \\
\mathcal{H}_\text{VB} \equiv \left(\mathcal{H}_\text{VV}+\mathcal{H}_\text{VC}\left(E-\mathcal{H}_\text{CC}\right)^{-1} \mathcal{H}_\text{CV}\right) 
\end{gather}
Using the expressions for $\mathcal{H}_\text{CC}$ etc.\ from Eq.\ \eqref{eq:ham_cbvb} above, and writing $\mathcal{H}_\text{CB}$ in the form given in Eq.\ \eqref{eq:ham_eff}, we arrive at the effective magnetic field given in Eq.\ \eqref{eq:Bsoc}, with the full expressions for $g_\parallel$ and $g_z$ in the conduction band given in Eq.\ \eqref{eq:beff_full} below. The expressions for the valence band are similar, with the replacements $\lambda_\text{VB} \rightarrow \lambda_\text{CB}$, $\lambda_\text{R} \rightarrow -\lambda_\text{R}$, and $\varepsilon_0 \rightarrow -\varepsilon_0$. When $\Delta$ is the dominant term, as is the case for heterostructures involving hBN, the effective magnetic field is well-approximated by the expressions in Eq.\ \eqref{eq:terms_BN}.

\begin{align}
g_\parallel(k) &= \frac{\lambda_\text{VB} \left( \Delta^2 - \lambda_\text{I}^2 - \lambda_\text{R}^2 + \left(\varepsilon_\text{PIA}^-\right)^2 \right) + 2\lambda_\text{VZ} \left( \lambda_\text{I}\varepsilon_\text{PIA}^- - \Delta\lambda_\text{R} \right) + 2\left( \varepsilon_0 + E \right) \left( \Delta\varepsilon_\text{PIA}^- - \lambda_\text{I}\lambda_\text{R} \right)}{\left( \varepsilon_0 + E \right)^2 - \lambda_\text{VZ}^2 - \lambda_\text{VB}^2} \nonumber \\
g_z(k) &= \frac{\lambda_\text{VZ} \left( \Delta^2 + \lambda_\text{I}^2 - \lambda_\text{R}^2 - \left(\varepsilon_\text{PIA}^-\right)^2 \right) + 2\lambda_\text{VB} \left( \lambda_\text{I}\varepsilon_\text{PIA}^- + \Delta\lambda_\text{R} \right) + 2\left( \varepsilon_0 + E \right) \left( \Delta\lambda_\text{I} + \lambda_\text{R}\varepsilon_\text{PIA}^- \right)}{\left( \varepsilon_0 + E \right)^2 - \lambda_\text{VZ}^2 - \lambda_\text{VB}^2}
\label{eq:beff_full}
\end{align}
\end{widetext}

\section{Density of States and Valence Charge Density}
\label{app:B}

In order to find out more about the origin and the magnitude of the proximity-induced SOC in a specific graphene/hXN heterostructure, we calculate the density of states (DOS) and the valence charge density in an energy window of $\pm 100$~meV around the Fermi level. For this purpose, we use a denser $k$-point grid of $42\times42\times1$ ($36\times36\times1$) for heterostructures with hBN and hGaN (hAlN).

%-----------------------------------------------------------------------------
\begin{figure}[htb]
 \includegraphics[width=0.98\columnwidth]{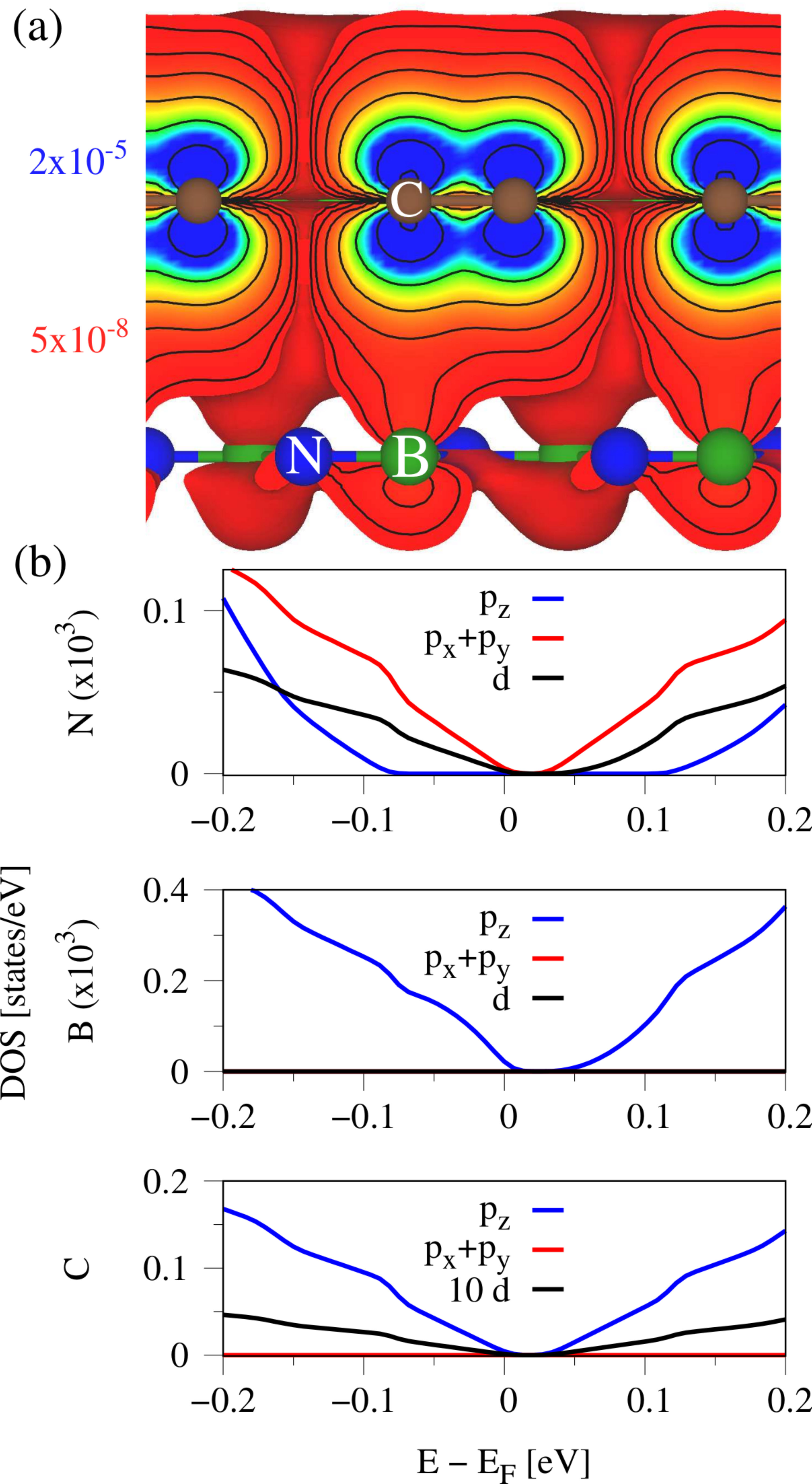}
 \caption{(Color online) (a) Calculated valence charge density of the graphene/hBN heterostructure considering only states in an energy window of $\pm 100$~meV around the Fermi level. The colors correspond to isovalues between $2\times10^{-5}$ (blue) and $5\times10^{-8}$ (red) e/\AA$^3$, while the isolines range from $1\times10^{-3}$ to $1\times10^{-7}$ e/\AA$^3$. (b) The atom and orbital resolved DOS.
 The DOS of B and N atoms is multiplied by a factor of $10^3$ for comparative reasons, while for C atoms only the $d$-orbital contribution is multiplied by a factor of 10.}
 \label{fig:hBN_DOS_valchargedens}
\end{figure}
%-----------------------------------------------------------------------------

In Fig. \ref{fig:hBN_DOS_valchargedens}, we show the calculated DOS and valence charge density for the graphene/hBN heterostructure. From the DOS, see Fig. \ref{fig:hBN_DOS_valchargedens}(b), we can see that mainly the $p_z$-orbitals from B contribute near the Dirac point, while additionally N $p_x+p_y$-orbitals are present. 
The calculated valence charge density, see Fig. \ref{fig:hBN_DOS_valchargedens}(a), clearly supports this picture, where a hybridization to B $p_z$-orbitals can be directly seen. 
The presented results are in agreement with recent calculations of the graphene/hBN system \cite{Zollner2019:PRB}, from where we also know that N and B atoms give a significant contribution to the intrinsic SOC parameters. 
The Rashba SOC should depend more on the deformation of C $p_z$-orbitals along the $z$-axis \cite{Gmitra2009:PRB} due to the hXN substrate. From the 
charge density plot, we cannot directly see such a deformation, hence the small Rashba SOC for the hBN case. 

In order to gain some more insights, we have also calculated the percentage atomic composition of the total DOS at energies of $\pm 100$~meV away from the Fermi level. 
At $100$~meV ($-100$~meV), the total DOS of the graphene/hBN structure is composed of
99.741\% C atoms, 0.189\% B atoms, and 0.069\% N atoms (99.622\% C atoms, 0.291\% B atoms, and 0.086\% N atoms).
Consequently, the valence band Dirac states are affected more strongly by the substrate than the conduction band Dirac states. This could be related to the fact that the $|\lambda_{\textrm{I}}^\textrm{B}|$ parameter, being proportional to the valence band splitting at K, deviates much stronger from the intrinsic SOC of bare graphene \cite{Gmitra2009:PRB} of about $12~\mu$eV, compared to the $|\lambda_{\textrm{I}}^\textrm{A}|$ parameter, being proportional to the conduction band splitting at K. 
In addition, one should keep in mind that the hBN valence band edge is much closer to the Dirac point than the conduction band edge, see Fig. \ref{Fig:bands}(a).
All in all, graphene is only
weakly perturbed by the hBN. 
Nevertheless, already such tiny substrate contributions strongly affect the SOC of the Dirac bands.

%-----------------------------------------------------------------------------
\begin{figure}[htb]
 \includegraphics[width=0.98\columnwidth]{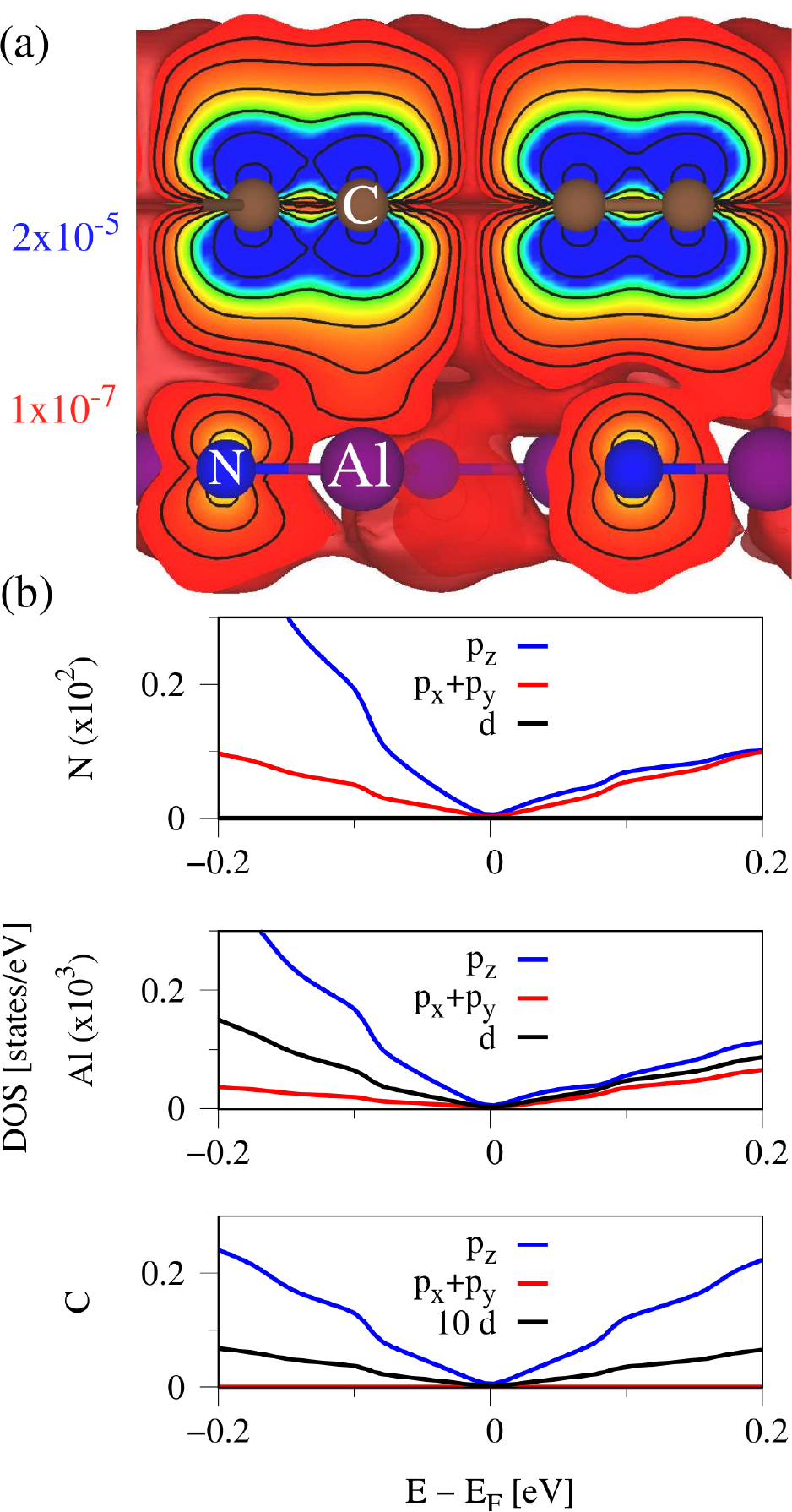}
 \caption{(Color online) (a) Calculated valence charge density of the graphene/hAlN heterostructure considering only states in an energy window of $\pm 100$~meV around the Fermi level. The colors correspond to isovalues between $2\times10^{-5}$ (blue) and $1\times10^{-7}$ (red) e/\AA$^3$, while the isolines range from $1\times10^{-3}$ to $1\times10^{-7}$ e/\AA$^3$. (b) The atom and orbital resolved DOS.
 The DOS of Al (N) atoms is multiplied by a factor of $10^3$ ($10^2$) for comparative reasons, while for C atoms only the $d$-orbital contribution is multiplied by a factor of 10.}
 \label{fig:hAlN_DOS_valchargedens}
\end{figure}
%-----------------------------------------------------------------------------

In Fig. \ref{fig:hAlN_DOS_valchargedens}, we show the calculated DOS and valence charge density for the graphene/hAlN heterostructure. From the DOS, see Fig. \ref{fig:hAlN_DOS_valchargedens}(b), we can see that mainly the $p_z$-orbitals from N and Al contribute for the valence Dirac bands. 
In contrast, for the conduction band side, also N $p_x+p_y$ give a strong contribution. 
The calculated valence charge density, see Fig. \ref{fig:hAlN_DOS_valchargedens}(a), again supports the DOS picture. 
Similarly to the hBN case, a clear deformation of the C $p_z$-orbitals along the $z$-axis is absent, which is consistent with the rather small Rashba SOC. 
Similar to before, we also look at the percentage atomic composition of the total DOS at $\pm 100$~meV away from the Fermi level. 
At $100$~meV ($-100$~meV), the total DOS of the graphene/hAlN structure is composed of
98.898\% C atoms, 0.083\% Al atoms, and 1.018\% N atoms (97.976\% C atoms, 0.164\% Al atoms, and 1.860\% N atoms).
In contrast to the hBN case, the total N contribution is enhanced by a factor of 10.
The asymmetry in the total hAlN DOS contribution for positive and negative energies is again related to the band structure, see Fig. \ref{Fig:bands}(b), where the hAlN valence band edge is only about 700~meV below the Dirac point.

%-----------------------------------------------------------------------------
\begin{figure}[htb]
 \includegraphics[width=0.98\columnwidth]{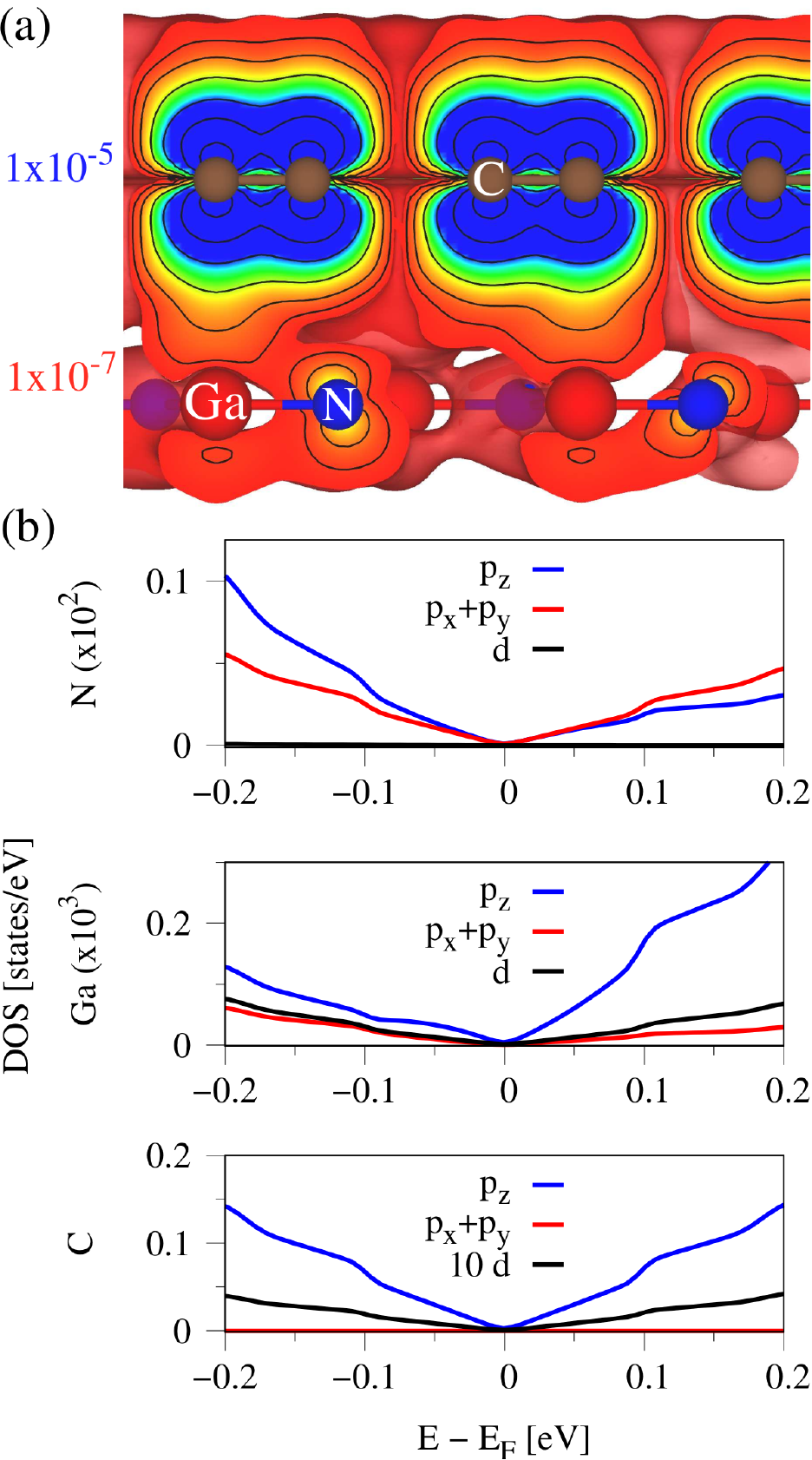}
 \caption{(Color online) (a) Calculated valence charge density of the graphene/hGaN heterostructure considering only states in an energy window of $\pm 100$~meV around the Fermi level. The colors correspond to isovalues between $1\times10^{-5}$ (blue) and $1\times10^{-7}$ (red) e/\AA$^3$, while the isolines range from $1\times10^{-3}$ to $1\times10^{-7}$ e/\AA$^3$. (b) The atom and orbital resolved DOS.
 The DOS of Ga (N) atoms is multiplied by a factor of $10^3$ ($10^2$) for comparative reasons, while for C atoms only the $d$-orbital contribution is multiplied by a factor of 10.}
 \label{fig:hGaN_DOS_valchargedens}
\end{figure}
%-----------------------------------------------------------------------------

In Fig. \ref{fig:hGaN_DOS_valchargedens}, we show the calculated DOS and valence charge density for the graphene/hGaN heterostructure. From the DOS, see Fig. \ref{fig:hGaN_DOS_valchargedens}(b), we find that mainly the $p$-orbitals from N contribute near the Dirac point, about 10-times more than for the hBN case but similar to the hAlN case. Moreover, there is some small contribution from Ga $p_z$-orbitals in the chosen energy window around the Dirac point.
The calculated valence charge density, see Fig. \ref{fig:hGaN_DOS_valchargedens}(a), clearly supports this picture, where a hybridization to N $p_z$-orbitals can be seen, which themselves couple to Ga $p$ orbitals. 
In addition, the C atoms which sit above the Ga atoms display a clear deformation of the $p_z$-orbitals along the $z$-axis (especially looking at the green colored isovalues). 
This deformation should be responsible for the rather large Rashba SOC in the hGaN case, compared to the other substrates. 
At $100$~meV ($-100$~meV), the total DOS of the graphene/hGaN structure is composed of
99.140\% C atoms, 0.261\% Ga atoms, and 0.599\% N atoms (98.944\% C atoms, 0.127\% Ga atoms, and 0.929\% N atoms).

\bibliography{paper}

\end{document}